\documentclass[11pt,a4paper]{article}

\catcode`\@=11
\@addtoreset{equation}{section}

\usepackage{amsbsy,amssymb,latexsym,amsfonts,amsmath}
\usepackage{mathrsfs}
\usepackage{graphicx}
\usepackage{hyperref}

\newcommand{\beq}{\begin{equation}}
\newcommand{\eeq}{\end{equation}}
\newcommand{\bea}{\begin{eqnarray}}
\newcommand{\eea}{\end{eqnarray}}

\newcommand{\CB}{{\mathcal B}}
\newcommand{\CC}{{\mathcal C}}

\newcommand{\CF}{{\mathcal F}}

\newcommand{\CN}{{\mathcal N}}
\newcommand{\CO}{{\mathcal O}}
\newcommand{\CP}{{\mathcal P}}

\newcommand{\ex}{\mathrm{e}}
\newcommand{\de}{\mathrm{d}}

\renewcommand\Re{{\mathrm{Re}}}

\newcommand\langled{\langle\!\langle}
\newcommand\rangled{\rangle\!\rangle}

\renewcommand{\thefootnote}{\fnsymbol{footnote}}

\begin{document}

%
%
\begin{titlepage}

\begin{flushright}
\normalsize
Imperial/TP/2014/KM/01
~~~~\\
\end{flushright}

\vspace{80pt}

\begin{center}
{\LARGE $\beta$-deformed Matrix Models and 2d/4d correspondence}\\
\end{center}

\vspace{25pt}

\begin{center}
{
Kazunobu Maruyoshi\footnote{k.maruyoshi@imperial.ac.uk} 
}\\
%
\vspace{15pt}
%
{\it Imperial Collage London, The Blackett Laboratory, Prince Concert Rd, London, SW7 2AZ, UK}\\
\end{center}
%
\vspace{20pt}
\begin{center}
Abstract\\
\end{center}
  We review the $\beta$-deformed matrix model approach to the correspondence between
  four-dimensional $\mathcal{N}=2$ gauge theories and two-dimensional conformal field theories.
  The $\beta$-deformed matrix model equipped with the log-type potential is obtained 
  as a free field (Dotsenko-Fateev) representation of the conformal block 
  of chiral conformal algebra in two dimensions, with the precise choice of integration contours. 
  After reviewing various matrix models related to the conformal field theories in two-dimensions,
  we study the large $N$ limit corresponding to turning off the Omega-background 
  $\epsilon_{1}, \epsilon_{2} \rightarrow 0$.
  We show that the large $N$ analysis produces the purely gauge theory results.   
  Furthermore we discuss the Nekrasov-Shatashvili limit ($\epsilon_{2} \rightarrow 0$) 
  by which we see the connection with the quantum integrable system.
  We then perform the explicit integration of the matrix model.
  With the precise choice of the contours 
  we see that this reproduces the expansion of the conformal block and also the Nekrasov partition function.
  This is a contribution to the special volume on the 2d/4d correspondence, edited by J.~Teschner.

\vfill

\setcounter{footnote}{0}
\renewcommand{\thefootnote}{\arabic{footnote}}

\end{titlepage}

\tableofcontents
\section{Introduction}
\label{sec:intro}
  
  Matrix models have played a crucial role in the studies of theoretical physics.
  It has turned out that these models compute quantum observables 
  or the partition function of quantum field theory \cite{Brezin:1977sv} 
  and two-dimensional gravity \cite{Ginsparg:1993is,DiFrancesco:1993nw} (see references therein).
  Rather recent examples are a one-matrix model which describes the low energy effective superpotential
  of four-dimensional $\CN=1$ supersymmetric gauge theory \cite{Dijkgraaf:2002dh}, 
  and the exact partition functions of supersymmetric gauge theories in various dimensions 
  \cite{Pestun:2007rz,Kapustin:2009kz} which are itself written as matrix models
  (Reviews can be found in \cite{P,H} in this volume).
  These have already shown the usefulness of the matrix model in theoretical physics.
  
  This paper reviews the matrix model introduced by Dijkgraaf and Vafa \cite{Dijkgraaf:2009pc} 
  which was proposed to capture the non-perturbative dynamics 
  of four-dimensional $\CN=2$ supersymmetric gauge theory and two-dimensional conformal field theory (CFT).
  This proposal is strongly related with the remarkable relation between 
  the Nekrasov partition function \cite{Nekrasov:2002qd} of four-dimensional $\CN=2$ supersymmetric gauge theory 
  and the conformal block of two-dimensional Liouville/Toda field theory found by \cite{Alday:2009aq}.
  (We refer to this relation as AGT relation [V:3].)
  The four-dimensional gauge theory is obtained by a partially twisted compactification of 
  the six-dimensional $(2,0)$ theory on a Riemann surface \cite{Gaiotto:2009we}, \cite{Ga}, 
  and the associated conformal block is defined on the same Riemann surface 
  where vertex operators are inserted at the punctures \cite{T}.
  
  The conformal block has several different representations.
  The one we focus here on is the Dotsenko-Fateev integral representation \cite{Dotsenko:1984nm,Dotsenko:1984ad},
  which will be interpreted as $\beta$-deformed matrix model.
  This integral representation has long been known, 
  but regarded as describing degenerate conformal blocks where
  the degenerate field insertion restricts the internal momenta to fixed values depending on the external momenta.
  However the recent proposal by \cite{Dijkgraaf:2009pc} is that it {\it does} describe the full conformal block.
  The point is the prescription of the contours of the integrations
  which divides integrals into sets of integral contours whose numbers are $N_{i}$
  (with $\sum N_{i} = N$ where $N$ is the size of the matrix.)
  In other words, in the large $N$ perspective, we fix the filling fractions when evaluating the matrix model.
  This gives additional degrees of freedom corresponding to the internal momenta.
  
  This matrix model plays an interesting role to bridge a gap between four-dimensional $\CN=2$ gauge theory
  on the $\Omega$ background and two-dimensional CFT.
  In addition to the correspondence with the CFT mentioned above, 
  this is because the matrix model has a standard expansion in $1/N$.
  The large $N$ limit in the matrix model corresponds to the $\epsilon_{1,2} \rightarrow 0$ limit
  on the gauge theory side.
  Therefore, the matrix model approach is suited for the $\epsilon$ expansion of the Nekrasov partition function.
  
  In section \ref{sec:DF}, we derive the $\beta$-deformed matrix model with the logarithmic potential 
  starting from the free scalar field correlator in the presence of background charge.
  The case of the Lie algebra-valued scalar field is described by the $\beta$-deformation 
  of the quiver matrix model \cite{Marshakov:1991gc,Kharchev:1992iv,Kostov:1992ie}.
  We further see that the similar integral representation can be obtained 
  for the correlator on a higher genus Riemann surface.
  These matrix models are proposed to be identified with the Nekrasov partition functions 
  of four-dimensional $\CN=2$ (UV) superconformal gauge theories and the conformal blocks.
  
  In section \ref{sec:limit} we analyze these matrix models,
  by taking the size of the matrix $N$ large.
  The leading part of the large $N$ expansion is studied by utilizing the so-called loop equation.
  We identify the spectral curve of the matrix model 
  with the Seiberg-Witten curve of the corresponding four-dimensional gauge theory
  in the form of \cite{Witten:1997sc,Gaiotto:2009we}.
  We then see evidence of the proposal by checking that the free energy at leading order reproduces the
  prepotential of the gauge theory.
  
  In section \ref{sec:NS} another interesting limit 
  which keeps one of the $\Omega$ deformation parameter $\epsilon_{1}$ finite
  while $\epsilon_{2} \rightarrow 0$ in the four-dimensional side will be analyzed.
  This limit was considered in \cite{Nekrasov:2009uh,Nekrasov:2009ui,Nekrasov:2009rc} 
  to relate the four-dimensional gauge theory on the $\Omega$ background 
  with the quantization of the integrable system.
  We will see that the $\beta$-deformation is crucial for the analysis, and that the matrix model 
  indeed captures the quantum integrable system.
    
  In section \ref{sec:finiteN}, we will perform a direct calculation of the partition function of the matrix model
  keeping all the parameters finite.
  We compare the explicit result of the direct integration with the Virasoro conformal block 
  and with the Nekrasov partition function.
   
  We conclude in section \ref{sec:conclusion} with a couple of discussions.
  In appendix \ref{sec:Selberg}, we present the Selberg integral formula and its generalization
  which will be used in the analysis in section \ref{sec:finiteN}.

\section{Integral representation of conformal block}
\label{sec:DF}
  In this section, we introduce the $\beta$-deformed matrix model as a free field representation
  of the conformal block, and the proposal \cite{Dijkgraaf:2009pc} 
  that the matrix model is related to the four-dimensional gauge theory.
  In section \ref{subsec:DF}, we see the simplest version of this proposal: 
  the $\beta$-deformed one-matrix model with the logarithmic-type 
  potential\footnote{The matrix model with a logarithmic potential was first studied by Penner \cite{Penner} 
                     related to the Eular characteristic of a Riemann surface.}
  obtained from the correlator
  of the single-scalar field theory on a sphere
  corresponds to the four-dimensional $\CN=2$ $SU(2)$ linear quiver gauge theory.
  In section \ref{subsec:quiver}, we will introduce the quiver matrix model corresponding to the gauge theory
  with higher rank gauge group.
  We will then generalize this to the one associated with a generic Riemann surface in section \ref{subsec:torus}.

\subsection{$\beta$-deformed matrix model}
\label{subsec:DF}
  In \cite{Alday:2009aq}, it was found that the conformal block on a sphere with $n$ punctures can be identified 
  with the Nekrasov partition function of $\CN=2$ $SU(2)^{n-3}$ superconformal linear quiver gauge theory.
  We will first review the integral representation of the conformal block, first introduced 
  by Dotsenko and Fateev \cite{Dotsenko:1984nm,Dotsenko:1984ad},
  and interpret it as a $\beta$-deformed matrix model \cite{Mehta, Eynard:2008mz}.
  (See \cite{Kostov:1999xi} for a review of the relation between the matrix model and the CFT.)
  We then state the conjecture among the matrix model, the Nekrasov partition function, and the conformal block.
  
  We start with the free scalar field $\phi(z)$ 
    \bea
    \phi(z)
     =     q + p \log z + \sum_{n\neq0} \frac{\alpha_{n}}{n} z^{-n}, 
    \eea
  with the following commutation relations
    \bea
    [ \alpha_{m}, \alpha_{n} ]
     =   - m \delta_{m+n,0},~~~~
    [ p, q  ]
     =   - 1.
    \eea
  Thus, the OPE of $\phi(z)$ is 
    \bea
    \phi(z) \phi(w) \sim - \log (z - w).
    \eea
  The energy-momentum tensor is given by 
  $T(z) = - \frac{1}{2} :\partial \phi(z) \partial \phi(z):$ with the central charge $1$.
  
  Let us introduce a background charge $Q=b + 1/b$ at the point at infinity by changing the energy-momentum tensor 
    \bea
    T(z)
     =  - \frac{1}{2} :\partial \phi(z) \partial \phi(z): + \frac{Q}{\sqrt{2}} :\partial^{2} \phi(z):
     =    \sum_{n \in \mathbb{Z}} \frac{L_{n}}{z^{n+2}}.
    \eea
  The central charge with this background is $c = 1 +6 Q^{2}$.
  
  The Fock vacuum is defined by
    \bea
    \alpha_{n} |0 \rangle
     =     0, ~~~
    \langle 0 | \alpha_{-n}
     =     0, ~~~~~
    {\rm for}~n\geq -1.
    \eea
  The energy-momentum tensor satisfies the Virasoro constraints
    \bea
    \langle L_{n} \rangle
     =     0, ~~~~~
    {\rm for}~n\geq -1.
    \eea
  
  Now we consider the correlator $\langle \prod_{k=0}^{n-1} V_{\alpha_{k}}(w_{k}) \rangle$,
  where the vertex operator is defined by $V_{\alpha}(z)=:e^{\sqrt{2} \alpha \phi(z)}:$
  with conformal dimension $\Delta_{\alpha}=\alpha(Q-\alpha)$.
  This is nonzero only if the momenta satisfy the condition $\sum_{k=1}^{n} \alpha_{k} = Q$.
  To relax the condition, let us consider the following operators
    \bea
    Q_{+}
     =     \int d \lambda :e^{\sqrt{2}b \phi(\lambda)} :, ~~~~
    Q_{-}
     =     \int d \lambda :e^{\sqrt{2}b^{-1} \phi(\lambda)} :.
    \eea
  Since the integrand of each operator has conformal dimension $1$, the screening operators are dimensionless.
  Therefore we can insert these operators into the correlator without changing the conformal property.
  The insertion however changes the momentum conservation condition,
  thus we refer these as screening operators.
  By inserting $N$ screening operators $Q_{+}$ in the correlator we define
    \bea
    \hat{Z}
     =     \left< Q_{+}^N ~
           \prod_{k=0}^{n-1} V_{\alpha_k}(w_k) \right>,
           \label{freefieldC0n}
    \eea
  The momentum conservation condition now relates the external momenta and the number of integrals as 
  $\sum_{k=0}^{n-1} \alpha_k + b N = Q$.
  This adds one more degree of freedom, $b N$, to the model.
  Nevertheless, it is important to note that the momenta $m_{k}$ (or $\alpha_{k}$) cannot be 
  completely arbitrary because $N$ is an integer.
  This point will be discussed in section \ref{sec:finiteN}.
  
  By evaluating the OPEs, it is easy to obtain
    \bea
    \hat{Z}
     =     C(m_k, w_k) Z
           \label{ZC0n}
    \eea
  where $Z$ is of the matrix model like form
    \bea
    Z
     =     \int \prod_{I=1}^N d \lambda_I \prod_{I < J} (\lambda_I - \lambda_J)^{-2b^2} 
           e^{-\frac{b}{g_s} \sum_I W(\lambda_I)}
     \equiv
           e^{F_{m}/g_s^2},
           \label{betamatrix}
    \eea
  with the following potential
    \bea
    W(z)
     =     \sum_{k=0}^{n-2} 2m_k \log (z - w_k), ~~~~
    C(m_k, w_k)
     =     \prod_{k < \ell \leq n-2} (w_k - w_\ell)^{- \frac{2 m_k m_\ell}{g_s^2}}.
           \label{potentialC0n}
    \eea
  We have introduced the parameter $g_s$ by defining $\alpha_k = \frac{m_k}{g_s}$.
  (We will use parameters $\alpha_{k}$ and $m_{k}$ interchangeably below.)
  We also have taken $w_{n-1} \rightarrow \infty$ by which the corresponding term in $W(z)$ disappeared.
  While the dependence on $m_{n-1}$ cannot be seen in the potential, 
  this is recovered by the momentum conservation condition
    \bea
    \sum_{k=0}^{n-1} m_k + b g_s N
     =    g_s Q.
           \label{momentumconservationC0n}
    \eea

  Note that the hermitian matrix model corresponds to the $b = i$ case
  because the first factor in the integrand is the familiar vandermonde determinant.
  Also the cases with $b=i/2$ and $2i$ correspond to an orthogonal matrix and a symplectic matrix respectively.
  However for generic choice of $b$, there is no such expression in terms of a matrix.
  This integral expression is known as $\beta$ ensemble or $\beta$-deformed matrix model
  with $\beta = - b^{2}$.
  
  It is useful to rewrite the $\beta$ deformed matrix model \eqref{betamatrix} as
    \bea
    Z
     =     \langle N | \exp \left( \frac{1}{2 \pi i \sqrt{2}g_{s}} \oint dw W(w) \partial \phi(w) \right) 
           Q_{+}^{N} | 0 \rangle,
           \label{z}
    \eea
  where we defined $\langle N | := \langle 0| e^{-\sqrt{2}bNq}$.
  Thus the insertion of (the derivative of) the scalar field $\phi$ in the correlator \eqref{z} is written as
    \bea
    \partial \phi(z)
     =  - \frac{W'(z)}{\sqrt{2} g_{s}} - b \sqrt{2} \sum_{I} \frac{1}{z - \lambda_{I}},
          ~~~
    \phi(z)
     =  - \frac{W(z)}{\sqrt{2} g_{s}} - b \sqrt{2} \log \prod_{I} (z - \lambda_{I}),
          \label{collective}
    \eea
  in the matrix model average $\langle \ldots \rangle$ defined by
    \bea
    \langle \CO \rangle
     =     \frac{1}{Z} \int \prod_{I=1}^N d \lambda_I \prod_{I < J} (\lambda_I - \lambda_J)^{-2b^2} 
           \CO ~e^{-\frac{b}{g_s} \sum_I W(\lambda_I)}.
           \label{average}
    \eea
  Note that a similar expression as \eqref{z} in terms of free fermions was presented in \cite{Nekrasov:2002qd}
  to express the instanton partition function of $\CN=2$ gauge theory.
      
\paragraph{Relation to conformal block}
  The proposal \cite{Dijkgraaf:2009pc} is 
  that the partition function of this $\beta$-deformed matrix model can be identified with 
  the Virasoro conformal block, and the Nekrasov partition function 
  of four-dimensional $\CN=2$ $SU(2)^{n-3}$ linear quiver gauge theory.
  The relation to the former is
    \bea
    Z_{0}^{-1} \hat{Z}(\alpha_{k}, N_{i}, b, w_{k})
     =     \CB(\alpha_{k}, \alpha^{int}_{p}, b, w_{k}),
    \eea
  where $Z_{0}$ is defined such that the $\hat{Z}$ is expanded in $w_{k}$ as $\hat{Z} = Z_{0} (1 + \CO(w_{k}))$. 
  Here $\CB$ is the Virasoro $n$-point conformal block on the sphere
  and defined such that $\CB = 1 + \CO(w_{k})$.
  We will review this in section \ref{subsec:Virasoro}.
  The momenta $\alpha_{k}$ are identified with the external momenta of the conformal block, as it should be.
  The parameters $b$ and $w_{k}$ are defined in the conformal block side in the same way as the free field theory. 
  Thus, the only nontrivial point is the identification of the internal momenta $\alpha^{int}_{p}$ ($p=1,\ldots,n-3$).
  
\begin{figure}[t]
    \centering
    \includegraphics[width=10cm]{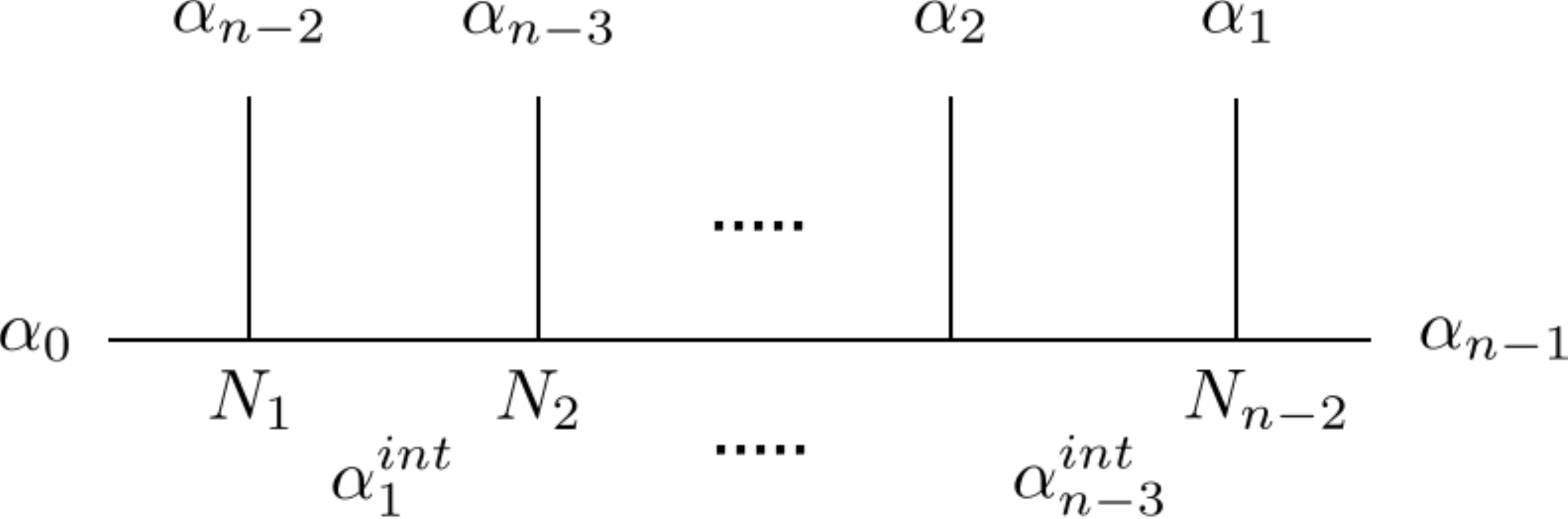}
    \caption{The $n$-point conformal block.
             The screening operators are inserted at each vertex to maintain the momentum conservation.}
    \label{fig:n-3}
    \end{figure}

  At the first sight there is no parameter corresponding to the internal momenta in the matrix model.
  However the prescription to identify them was established 
  by \cite{Mironov:2010zs,Itoyama:2010ki,Mironov:2010ym,Cheng:2010yw}:  
  as we will see in section \ref{subsec:Virasoro}, the conformal block can be computed from the three-point functions,
  denoted by the trivalent vertices, and the propagators, denoted by the lines connecting the vertices,
  as in figure \ref{fig:n-3}.
  The idea is that there are $N_{i}$ screening operators inserted at each vertex, with $\sum_{i=1}^{n-2} N_{i}=N$,
  where the momentum conservation is satisfied as
    \bea
    \alpha^{int}_{1}
    &=&    \alpha_{0} + \alpha_{n-2} + bN_{1},~~~~
    \alpha^{int}_{2}
     =     \alpha_{1}^{int} + \alpha_{n-3} + bN_{2}, ~~\ldots,
           \nonumber \\
    \alpha^{int}_{n-3}
    &=&    \alpha^{int}_{n-4} + \alpha_{2} + bN_{n-3}
     =   - \alpha_{1}-\alpha_{n-1}-bN_{n-2} +Q,
           \label{momcon}
    \eea
  In the last equality we used the momentum conservation \eqref{momentumconservationC0n}.
  This means that in the integral representation we have 
  $n-2$ sets of integrals, each number of the integrals is $N_{i}$.

  The precise choice of the integration contours will be seen in section \ref{sec:finiteN}.
  Here let us see a rationale of this identification by considering the large $N$ limit shortly.
  The critical points of the eigenvalues $\lambda_{I}$ are obtained from the equations of motion
    \bea
    \sum_{k = 0}^{n-2} \frac{m_k}{\lambda_I - w_{k}}
         + b g_s \sum_{J (\neq I)} \frac{1}{\lambda_I - \lambda_J}
     =     0.
           \label{eom}
    \eea
  Focusing on the first term, when the parameters are generic enough there are $n-2$ critical points.
  Let $N_{i}$ be the number of the matrix eigenvalues which are at the $i$-th critical point.
  These critical points are diffused to form line segments by the second term.
  The integrals are defined such that they include these segments.
  Now we introduce the filling fractions $\nu_{i} = b g_{s} N_{i}$,
  and consider the matrix model by fixing these values in the large $N$ limit.
  Because of the momentum conservation, we have $n-3$ independent degrees of freedom.

\paragraph{Relation to Nekrasov partition function}
  The relation to the Nekrasov partition function is as follows:
    \bea
    Z_{U(1)} Z_{0}^{-1} \hat{Z}(\alpha_{k}, N_{i}, b, w_{k})
     =      Z_{{\rm Nek}}(m_{k}, a_{p}, \epsilon_{1}, \epsilon_{2}, q_{p}),
            \label{Nek}
    \eea
  under the following identification of the parameters.
  We choose three insertion points as $w_0 = 0$, $w_1 = 1$ and $w_{n-1} = \infty$.
  The remaining parameters are identified with the gauge theory coupling constants 
  $q_p =e^{2 \pi i \tau_p}$ ($p=1, \ldots, n-3$) as follows:
    \bea
    w_2
     =     q_1, ~~
    w_3
     =     q_1 q_2, ~~ \ldots, ~~
    w_{n-2}
     =     q_1 q_2 \cdots q_{n-3}.
           \label{moduliidentificationC0n}
    \eea
  We denote the gauge group whose gauge coupling constant is $q_{p}$ as $SU(2)_{p}$.
  Let $\mu^{L}_{a}$, $\mu^{L}_{b}$ and $\mu^{R}_{a}$, $\mu^{R}_{b}$ be the mass parameters of 
  hypermultiplets in the fundamental representation of the $SU(2)_{1}$
  and those of the $SU(2)_{n-3}$ respectively.
  Let also $\mu_{i}$ ($i=1,\ldots,n-4$) be the mass parameter of the hypermultiplet
  in the $(\bf{2},\bar{2})$ representation of $SU(2)_{i} \times SU(2)_{i+1}$.
  Then the mass parameters and the external momenta are identified as
    \bea
    m_{0}
    &=&    \frac{\mu^{L}_{a} - \mu^{L}_{b}}{2} + \frac{g_{s} Q}{2},~~~
    m_{n-2}
     =     \frac{\mu^{L}_{a} + \mu^{L}_{b}}{2},
           \nonumber \\
    m_{n-1}
    &=&    \frac{\mu^{R}_{a} - \mu^{R}_{b}}{2} + \frac{g_{s} Q}{2},~~~
    m_{1}
     =     \frac{\mu^{R}_{a} + \mu^{R}_{b}}{2},~~~
    m_{n-2-i}
     =     \mu_{i}
    \eea
  The identification of the parameter $b$ with the $\Omega$-deformation parameters is given by 
    \bea
    \epsilon_1
     =     b g_s, ~~~
    \epsilon_2
     =     \frac{g_s}{b}.
           \label{epsilonrelation}
    \eea
  Note that the case $b=i$ corresponds to the self-dual background $\epsilon_1 = - \epsilon_2$. 
  Finally, the vacuum expectation values $a_{i}$ of the scalar fields in the $SU(2)_{i}$ vector multiplets are 
  identified as
    \bea
    a_{p} - \mu^{L}_{a} - \sum_{q=1}^{p-1} \mu_{q}
     =     \sum_{q=1}^{p} b N_{q},
    \eea
  for $p=1,\dots,n-3$.
  By using the momentum conservation, $a_{n-3}$ can also be written as $a_{n-3} + \mu^{R}_{a} = - b N_{n-2}$.
  
  The first factor in the right hand side of \eqref{Nek} is the so-called $U(1)$ factor corresponding 
  to the $U(1)$ part of the gauge theory, which is, e.g., given by
    \bea
    Z_{U(1)}
     =     (1-q)^{2\alpha_{1} \alpha_{2}},
    \eea
  for the $n=4$ case\footnote{This is slightly different from the one in \cite{Alday:2009aq}.
                              This is because we consider the Nekrasov partition function 
                              where the hypermultiplets are in the fundamental representation
                              of the gauge group.
                              Changing the representation to the anti-fundamental one leads to
                              $\alpha_{3} \rightarrow Q - \alpha_{3}$ in this case,
                              then we recover the factor in \cite{Alday:2009aq}}.

\subsection{Quiver matrix model and higher rank gauge theory}
\label{subsec:quiver}
  In this section, we briefly review the $\beta$-deformation of the ADE quiver matrix model 
  \cite{Marshakov:1991gc,Kharchev:1992iv,Kostov:1992ie,Itoyama:2009sc}.
  We then see that the matrix model can be obtained from the CFT of a free chiral boson
  valued in Lie algebra.
  A review of the undeformed quiver matrix model can be found in \cite{Chiantese:2003qb}.
  
  Let $\mathfrak{g}$ be a finite dimensional Lie algebra of ADE type with rank $r$,
  $\mathfrak{h}$ the Cartan subalgebra of $\mathfrak{g}$, and $\mathfrak{h}^*$ its dual. 
  We denote the natural pairings between $\mathfrak{h}$ and $\mathfrak{h}^*$ by $\langle \cdot, \cdot \rangle$:
    \beq
    \alpha(h)
     =     \langle \alpha, h \rangle, \qquad \alpha \in \mathfrak{h}^*, \ \ h \in \mathfrak{h}.
    \eeq
  Let $\alpha_a \in \mathfrak{h}^*$ $(a=1,2,\dotsc, r)$ be simple roots of $\mathfrak{g}$
  and $( \cdot, \cdot)$ is the inner product on $\mathfrak{h}^*$.
  Our normalization is chosen as $(\alpha_a, \alpha_a)=2$.
  The fundamental weights are denoted by $\Lambda^a$ $(a=1,2,\dotsc, r)$
    \beq
    ( \Lambda^a, \alpha_b^{\vee} )
     =     \delta^a_b, \qquad
    \alpha_a^{\vee}
     =     \frac{2 \alpha_a}{(\alpha_a, \alpha_a)}.
    \eeq
  
  In the Dynkin diagram of $\mathfrak{g}$ we associate $N_a \times N_a$ Hermitian matrices $M_a$
  with vertices $a$ for simple roots $\alpha_a$, 
  and complex $N_a \times N_b$ matrices $Q_{ab}$ and their Hermitian conjugate $Q_{ba} = Q_{ab}^{\dag}$ 
  with links connecting vertices $a$ and $b$. 
  We label links of the Dynkin diagram by pairs of nodes $(a,b)$ 
  with an ordering $a<b$.
  Let $\mathcal{E}$ and $\mathcal{A}$ be the set of ``edges'' $(a,b)$ 
  (with $a<b$) and the set of ``arrows'' $(a,b)$ respectively:
    \bea
    \mathcal{E}
    &=&   \{ (a,b) \, | \, 1 \leq a<b \leq r, \ (\alpha_a, \alpha_b) = -1 \}, 
           \nonumber \\
    \mathcal{A}
    &=&   \{ (a, b) \, | \, 1 \leq a,b \leq r, \ (\alpha_a, \alpha_b)= - 1 \}.
    \eea
  The partition function of the quiver matrix model 
  associated with $\mathfrak{g}$ is given by
    \beq
    Z
     =     \int \prod_{a=1}^r [ \de M_a ] \prod_{(a,b) \in \mathcal{A} } [ \de Q_{ab} ]
           \exp\left( \frac{1}{g_s} W(M, Q) \right),
    \eeq
  where
    \beq
    W(M, Q)
     =     i \sum_{(a,b) \in \mathcal{A}} s_{ab} \mathrm{Tr} \, Q_{ba} M_a Q_{ab}
         - i \sum_{a=1}^r \mathrm{Tr}\, W_a( M_a),
    \eeq
  with real constants $s_{ab}$ obeying the conditions $s_{ab} = - s_{ba}$. 
  Note that
    \beq
    \prod_{(a,b) \in \mathcal{A}}[ \de Q_{ab} ]
     =     \prod_{(a,b) \in \mathcal{E}} [ \de Q_{ba} \de Q_{ab} ],
    \eeq
    \beq
    \sum_{(a,b) \in \mathcal{A}} s_{ab} \mathrm{Tr} \, Q_{ba} M_a Q_{ab}
     =     \sum_{(a,b) \in \mathcal{E}} s_{ab} \bigl( \mathrm{Tr} \, Q_{ba} M_a Q_{ab} - \mathrm{Tr}\, 
           Q_{ab} M_b Q_{ba} \bigr).
    \eeq
  The integration measures $[\de M_a ]$ and $[ \de Q_{ba} \de Q_{ab} ]$ are defined 
  by using the metrics $\mathrm{Tr} ( \de M_a)^2$ and $\mathrm{Tr}( \de Q_{ba} \de Q_{ab})$ respectively.

  Integrations over $Q_{ab}$ are easily performed:
    \beq
    \label{GF}
    \int [ \de Q_{ba} \de Q_{ab} ] \exp\left( \frac{i s_{ab}}{g_s} 
    \bigl( \mathrm{Tr} \, Q_{ba} M_a Q_{ab} - \mathrm{Tr}\, Q_{ab} M_b Q_{ba} \bigr) \right)
     =     \det\bigl( M_a \otimes 1_{N_b} - 1_{N_a} \otimes M_b^T \bigr)^{-1},
    \eeq
  where $1_n$ is the $n \times n$ identity matrix and $T$ denotes transposition. For simplicity 
  we have chosen the normalization of the measure $[\de Q_{ba} \de Q_{ab}]$ to set the proportional constant 
  in the right hand side of \eqref{GF} to be unity.
  Now the integrand depends only on the eigenvalues of $r$ Hermitian matrices $M_a$.
  Let us denote them by $\lambda^{(a)}_I$ ($a=1,2,\dotsc, r$ and $I=1,2,\dotsc, N_a$).
  The partition function of the quiver matrix model reduces to the form of integrations over the eigenvalues of $M_a$ 
    \beq 
    \label{part}
    Z
     =     \int \prod_{a=1}^r \left\{ \prod_{I=1}^{N_a} \de \lambda^{(a)}_I \right\}\, 
           \Delta_{\mathfrak{g}}(\lambda) 
           \exp\left( - \frac{i}{g_{s}} \sum_{a=1}^r \sum_{I=1}^{N_a} W_a(\lambda^{(a)}_I) \right),
    \eeq
  where $W_a$ is a potential and
    \beq
    \Delta_{\mathfrak{g}}(\lambda)
     =     \prod_{a=1}^r \prod_{1 \leq I < J \leq N_a} ( \lambda_I^{(a)} - \lambda_J^{(a)})^2 \prod_{1 \leq a< b \leq r}
           \prod_{I=1}^{N_a} \prod_{J=1}^{N_b} ( \lambda^{(a)}_I - \lambda^{(b)}_J )^{(\alpha_a, \alpha_b)}.
    \eeq
  We then define the $\beta$ deformation of the above quiver matrix model (with $\beta = -b^2$) by
    \beq
    Z
     =     \int \prod_{a=1}^r \left\{ \prod_{I=1}^{N_a} \de \lambda^{(a)}_I \right\}
           \Bigl(\Delta_{\mathfrak{g}}(\lambda) \Bigr)^{-b^2} 
           \exp\left( - \frac{b}{g_s} \sum_{a=1}^r \sum_{I=1}^{N_a} W_a(\lambda^{(a)}_I) \right).
           \label{partb}
    \eeq
  At $b = i$, it reduces to the original quiver matrix model \eqref{part}.

  The partition function \eqref{partb} can be rewritten in terms of CFT operators.
  Let $\phi(z)$ be $\mathfrak{h}$-valued massless chiral field and $\phi_a(z):= \langle \alpha_a, \phi(z) \rangle$. 
  Their correlators are given by
    \beq
    \phi_a(z) \phi_b(w)
     \sim 
         - ( \alpha_a, \alpha_b ) \log(z-w), \qquad
           a,b=1,2,\dotsc, r.
    \eeq
  The modes
    \beq
    \phi(z)
     =     q + p \log z + \sum_{n \neq 0} \frac{a_n}{n} z^{-n} \in \mathfrak{h}
    \eeq
  obey the commutation relations
    \beq
    [ \langle \alpha, a_n \rangle, \langle \beta, a_m \rangle ]
     =   - n \delta_{n+m,0} ( \alpha, \beta), \qquad
    [ \langle \alpha, p \rangle, \langle \beta, q \rangle]
     =   - i ( \alpha, \beta), \qquad \alpha, \beta \in \mathfrak{h}^*.
    \eeq
  The Fock vacuum is given by
    \beq
    \alpha(a_n)|0 \rangle = 0, \qquad
    \langle 0 | \alpha(a_{-n}) = 0, \qquad
     n \geq 0, \qquad
    \alpha \in \mathfrak{h}^*.
    \eeq
  Let
    \beq
    \langle \{ N_a \} |
    :=     \langle 0| \exp\left( - b \sum_{a=1}^r N_a \alpha_a( \phi_0) \right).
    \eeq
  It is convenient to introduce the $\mathfrak{h}^*$-valued potential $W(z)$ by
    \beq
    W(z)
    :=     \sum_{a=1}^r W_a(z) \Lambda^a \in \mathfrak{h}^*.
    \eeq
  Note that $W_a(z) = ( \alpha^{\vee}_a, W(z) )$.

  As in the previous subsection, we put the background charge $Q = b + 1/b$ 
  which leads to the energy-momentum tensor
    \beq
    T(z)
     =   - \frac{1}{2} : \mathcal{K}\bigl( \partial \phi(z), \partial \phi(z) \bigr):
         + Q \langle \rho, \partial^2 \phi(z) \rangle,
    \eeq
  where $\mathcal{K}$ is the Killing form
  and $\rho$ is the Weyl vector of $\mathfrak{g}$, half the sum of the positive roots.
  Let $H^i$ ($i=1,2,\dotsc, r$) be an orthonormal basis of the Cartan subalgebra $\mathfrak{h}$ 
  with respect to the Killing form: $\mathcal{K}(H^i, H^j) = \delta^{ij}$. 
  In this basis, 
  the components of the $\mathfrak{h}$-valued chiral boson are just $r$ independent free chiral bosons: 
    \beq
    \phi(z)
     =      \sum_{i=1}^r H^i \phi_i(z), \qquad
    \phi_i(z) \phi_j(w) \sim - \delta_{ij} \log(z-w),
    \eeq
  and the energy-momentum tensor in this basis is given by
    \beq
    T(z)
     =   - \frac{1}{2} \sum_{i=1}^r : \bigl( \partial \phi_i(z) \bigr)^2:
         + Q \sum_{i=1}^r \rho^i \partial^2 \phi_i(z).
    \eeq
  The central charge is given by
    \beq
    \label{ccADE}
    c
     =     r + 12 Q^2 ( \rho, \rho) = r \Bigl\{ 1 + h ( h+1) Q^2 \Bigr\}.
    \eeq
  Here $h$ is the Coxeter number of the simply-laced Lie algebra $\mathfrak{g}$ whose rank is $r$. 
  Explicitly, $h_{A_{n-1}} = n$ (with $r=n-1$), $h_{D_r} =2r-2$, $h_{E_6}=12$, $h_{E_7}=18$ and $h_{E_8}= 30$.

  Note that for a root $\alpha$, $[H^i, E_{\alpha}] = \alpha^i E_{\alpha}$ 
  with $\alpha^i = \alpha(H^i) = \langle \alpha, H^i \rangle$.
  Then, the bosons $\phi_a(z)$ associated with the simple roots $\alpha_a$ are expressed in this basis as follows:
    \beq
    \phi_a(z)
     =     \langle \alpha_a, \phi(z) \rangle
     =     \sum_{i=1}^r \alpha_a^i \phi_i(z) \equiv \alpha_a \cdot \phi(z), 
           \qquad a=1,2,\dotsc, r.
    \eeq
  For roots $\alpha$ and $\beta$, the inner product on the root space is expressed in their components as
  $( \alpha, \beta) = \sum_{i=1}^r \alpha^i \beta^i$. 
  Here $\alpha^i = \alpha(H^i)$ and $\beta^i = \beta(H^i)$.
  
  Let us now consider the four-point correlator of this theory.
  The vertex operator is defined by 
    \bea
    \label{VOdet}
    V_{\hat{\mu}}(z)
     =     : \ex^{\left< \hat{\mu}, \phi(z) \right>} :,
    \eea
  where $\hat{\mu} \in \mathfrak{h}^{*}$.
  As in the one-matrix case, we introduce the screening operators associated with the simple roots are defined by
    \beq
    Q_a
    :=     \int \de z \, : \ex^{b \phi_a(z)}:, \qquad a=1,2,\dotsc, r.
    \eeq
  We define the chiral four-point correlation function 
    \begin{equation}
    \hat{Z}
     =     \left< :\prod_{k=0}^3 \ex^{\langle \hat{\mu}_k , \phi(w_k) \rangle}: Q_1^{N_1} Q_2^{N_2} \dotsm Q_{r}^{N_{r}}    
           \right>.
    \end{equation}
  For later convenience, we set $m_k:= g_s \hat{\mu}_k$ $(k=0,1,2,3)$.
  The momentum conservation condition is required
    \bea
    \label{momentumconservation}
    \sum_{k=0}^3 m_k + \sum_{a=1}^{r} b g_{s} N_a \alpha_a = 0.
    \eea
  Using this four-point function, we define the partition function of the $\beta$ deformed quiver matrix model 
  by sending $w_{3} \rightarrow \infty$ \eqref{partb} with the potential $W_a(z)$:
    \bea
    W_a(z)
     =     \sum_{k = 0}^2 \left( m_k, \alpha_a \right)
           \log (w_k - z).
           \label{matrixactionn}
    \eea
  We will set $w_0 = 0$, $w_1 = 1$ and $w_2 = q$.
  Using these definitions, the partition function \eqref{part} can be written as follows
    \beq
    Z
     =      \langle \{ N_a \} | \, \exp\left( \frac{1}{2\pi i g_s} \oint_{\infty}
            \de z \, \langle W(z), \partial \phi(z) \rangle \right) 
            ( Q_1)^{N_1} \dotsm (Q_r)^{N_r} \, | 0 \rangle.
    \eeq

\subsection{Higher genus case}
\label{subsec:torus}
  A generalization of the matrix model to a higher genus Riemann surface has also been considered 
  in \cite{Dijkgraaf:2009pc}.
  The integral representation is basically obtained by changing the two-point function of the free field 
  on a sphere to the one on a Riemann surface, which can be written in terms of the prime form,
  and by adding a term to the action 
  which is the integral of the holomorphic differentials on the Riemann surface.
  For the conformal block on a torus with $n$ punctures, for instance, 
  the two-point function is proportional to the theta function 
  and the integral representation is given by \cite{Dijkgraaf:2009pc,Maruyoshi:2010pw}
    \bea
    Z
    &=& 
           \int \prod_{I=1}^N d \lambda_{I} 
           \prod_{1 \le I < J \le N} \theta_1 (\lambda_{I}-\lambda_{J})^{-2b^2}
           \exp \left( - \frac{b}{g_s} \sum_{I=1}^N W(\lambda_{I}) \right),
           \label{holo}
    \eea
  where $\theta_1(z) = 2 q^{1/8} \sin z \prod_{n=1}^\infty (1 - q^n) (1 - 2q^n \cos 2 z + q^{2n})$,
  $q = \exp(2 \pi i \tau)$, and
    \bea
    W(z)
     =     \sum_{k=1}^n 2 m_k \log \theta_1 (z-w_k) + 4 \pi i a z.
           \label{c_action}
    \eea
  The last term in $W(z)$ is the integral of the holomorphic differential on the torus, $dz$, as mentioned above.
  Since the factor $\prod_{1 \le I < J \le N} \theta_1 (\lambda_{I}-\lambda_{J})^{-2b^2}$ can be regarded
  as the generalization of the Vandermonde determinant,
  we refer to the integral (\ref{holo}) as ``generalized matrix model''.
  In \cite{Dijkgraaf:2009pc}, the potential (\ref{c_action}) of the generalized matrix model was 
  expected from the geometrical argument of topological string theory.
  
  In the following, we explain how the generalized matrix model is obtained 
  from the full Liouville correlation function 
  \cite{Maruyoshi:2010pw} for the torus case and \cite{Bonelli:2010gk} for the generic Riemann surface, 
  based on the perturbative argument of \cite{Goulian:1990qr}.
  This method is different from the one seen in the previous subsection, 
  although the both use the free field formalism. 

  The $n$-point function of the Liouville theory on a genus $g$ Riemann surface $\CC_{g}$
  is formally given by the following path integral 
    \bea
    A 
     \equiv \left< \prod_{k=1}^n e^{2 \alpha_k\phi(w_k, \bar{w}_k)} \right>_{{\rm Liouville \,\, on}\,\,\CC_{g}}
     =     \int {\mathcal{D}} \phi (z,\bar{z}) e^{-S[\phi]} \prod_{k=1}^n e^{2\alpha_k \phi (w_k,\bar{w}_k)},
    \eea
  where the Liouville action is given by 
    \bea
    S[\phi] 
     =     \frac{1}{4 \pi}\int d^2 z \sqrt{g} ( \partial_a \phi \partial^a \phi + Q R \phi + 4 \pi\mu e^{2b \phi}).
    \eea
  Here $R$ is Ricci scalar and $\mu$ is a constant.
  We divide the Liouville field into the zero mode and the fluctuation
  $\phi (z,\bar{z}) = \phi_0 + \tilde{\phi} (z,\bar{z})$.
  By integrating over $\phi_0$, we obtain
    \bea
    A 
     &=&    \frac{\mu^{N} \Gamma(-N)}{2b}
            \int {\mathcal{D}} \tilde{\phi} (z,\bar{z}) e^{-S_0[\tilde{\phi}]} 
            e^{-\frac{Q}{4 \pi} \int d^{2} z R \tilde{\phi}}
            \left( \int d^2 z \,\, e^{2b \tilde{\phi} (z,\bar{z})} \right)^{N}
            \prod_{k=1}^n e^{2\alpha_k \tilde{\phi} (w_k,\bar{w_k})} ,
    \eea
  where 
    \bea
    N
     =   - \sum_{k=1}^{n} \frac{\alpha_k}{b} + \frac{Q}{b}(1 - g),
    \eea
  and $S_0$ is the free scalar field action.
  When $N \in \mathbb{Z}_{\geq 0}$, the correlator diverges due to the factor $\Gamma(-N)$.
  The residues at these poles $A_{N}$ are evaluated in the perturbation theory in $b$ around the free field action:
    \bea
    A_{N}
    &=&    \frac{(-\mu)^{N}}{2b N!}
           \int \prod_{I=1}^N d^2 z_I \left\langle e^{-\frac{Q}{4 \pi} \int d^{2} z R \tilde{\phi}} 
           \prod_{I=1}^N :e^{2 b \phi (z_I, \bar{z}_I)}: 
           \,\, \prod_{k=1}^n : e^{2 \alpha_k \phi (w_k, \bar{w}_k)} : 
           \right\rangle _{{\rm free \,\, on} \,\, \CC_{g}}.
           \label{A1}
    \eea
  That $N$ is integer ensures the momentum conservation in the free theory.
  
  Now let us focus on the torus case which simplifies the expression.
  The $\ell$-point function of the free theory on a torus is written in terms of the factorized expression
  by introducing an additional integral as \cite{Verlinde:1987sd,Dijkgraaf:1987vp,D'Hoker:1988ta}
    \bea
    &&\left\langle \prod_{i=1}^{\ell} :e^{i k_i \phi (z_i, \bar{z}_i)}: \right\rangle 
    _{{\rm free \,\, on} \,\, T^2}
    \cr
    &&=     2 i |\eta (\tau)|^{-2} \delta ( \sum _i k_i )  
            \int^{\infty}_{-\infty} da \left| \left( \prod_{i<j} 
           \left( \frac{\theta_1 \left( z_{ij} | \tau \right)}{\eta(\tau)^{3}} \right) 
           ^{\frac{k_ik_j}{2}} \right) q^{a^2}
           \exp \left( - 2 \pi i \sum_{j=1}^{\ell} k_j z_j a \right)
           \right| ^2 ,
           \label{factorized}
    \eea
  where $z_{ij} \equiv z_i-z_j$, $\tau$ is the moduli of the torus and $q=\exp(2 \pi i \tau)$.
  By using the explicit expression (\ref{factorized}), 
  we find that the $n$-point function $A_{N}$ of the Liouville theory reduces to the following integral 
    \bea
    & &
    A_{N}
     =     C (\tau,m_k,b) \prod_{1\le k<l \le n} 
           \left| \theta_1 ( w_{kl} )\right| ^{-4m_k m_l}
           \int^{i\infty}_{-i\infty} da |q|^{-2a^2} \int_{T^2} \prod_{I=1}^N d^2 z_I 
           \label{Amp_ctdt} \\
    & &  ~~~~~ \times  \left| \exp \left[ 
           - 2b \sum_{I=1}^N \sum_{k=1}^n m_k \log \theta_1 \left( z_{I}-w_{k} \right)
           - 2b^2 \sum_{I<J} \log \theta_1 ( z_{IJ} ) 
           - 4 \pi i b a  \sum_{I=1}^N z_I
           \right] \right|^2,
           \nonumber 
    \eea
  where $w_{kl} \equiv w_k-w_l$, and we have chosen the insertion points $w_{k}$ such that 
  they satisfy $\sum_{k} m_{k} w_{k} = 0$.
  The factor $C(\tau,m_k,b)$ in front of the $z$ integral is irrelevant for the analysis below.
  
  The discussion above is valid even for finite $N$.
  However, it is not straightforward to divide the integral over the torus into the product of
  the holomorphic and the anti-holomorphic pieces for generic $N$.
  In order to proceed, we evaluate the integral (\ref{Amp_ctdt}) in the large $N$ limit.
  We see that all the three terms in the exponent in (\ref{Amp_ctdt}) are ${\cal O}(N^2)$.
  Thus, the integral (\ref{Amp_ctdt})
  is evaluated at the critical points of the exponent of the integrand.
  The conditions for the criticality of the exponent are factorized into
  holomorphic equations and anti-holomorphic equations,
  which indicates that the integral over the torus in (\ref{Amp_ctdt}) can be replaced 
  by the product of the holomorphic and the anti-holomorphic integrals in the large $N$ limit.
  Thus we define the holomorphic part of the correlation function as in \eqref{holo}
  after introduction of $g_{s}$ by $\alpha_{k}=m_{k}/g_{s}$.

\paragraph{Relation to conformal block and gauge theory}
  We propose that this generalized matrix model \eqref{holo} reproduces the full conformal block
  on the punctured torus (not only in the large $N$ limit), 
  and also the Nekrasov partition function of the $\CN=2$ elliptic $SU(2)^{n}$ quiver gauge theory
  which is obtained from two M5-branes on the same torus.
  
  Let us shortly see the relation of the parameters in the conformal block and the generalized matrix model.
  In the toric conformal block with $n$ punctures, we have $n$ external and $n$ internal momenta,
  giving $2n$ parameters in total.
  The parameters $m_k$ are directly identified with the external momenta.
  Then, the potential \eqref{c_action} has $n$ critical points for each variable $z_I$, 
  assuming that the parameters $m_k$ are generic.
  Similar to the case in subsection \ref{subsec:DF}, we expect that the $n$ critical points are ``diffused'' 
  to form line segments due to the ``determinant'' factor.
  Then, the partition function is labelled by the filling fractions $\nu_i = b g_s N_i$,
  in which $N_i$ out of $N$ variables $z_I$ take the value on the $i$-th line segment.
  Due to the momentum conservation condition the sum of all $\nu_{i}$ is not independent degree of freedom.
  Thus we have $n-1$ independent filling fractions.
  These and the parameter $a$ in the potential are mapped to the internal momenta.
  (See \cite{Mironov:2010su} for the precise identification in the $n=1$ case.)
  
  The relation to the gauge theory is stated as follows:
  the gauge theory coupling constants $q_p = e^{2 \pi i \tau_p}$ ($p=1, \ldots, n$) are identified 
  with the moduli of the torus as
    \bea
    e^{2 \pi i w_{k}}
     =     \prod_{p=k}^{n-1} q_{p}, ~~~
    q 
     \equiv 
           e^{2 \pi i \tau}
     =     \prod_{p=1}^{n} q_p.
    \eea
  The parameters $m_k$ are directly identified with the mass parameters of the bifundamentals.
  The filling fractions and the parameters in the potential are mapped to the vevs $a_{p}$
  of the scalars in the vector multiplets.

\paragraph{$g>1$ case}
  Finally, let us quickly consider the case of the genus $g$ Riemann surface with $n$ puncture.
  As stated above the two-point function is written in terms of the prime form,
  and the generalized matrix model is the one in \eqref{c_action} where the theta function is 
  replaced by the prime form and the last term in the potential is the integral of the holomorphic differential,
  with some additional terms.
  The precise form is presented in \cite{Bonelli:2010gk}. 
  The parameters are identified as follows \cite{Mironov:2010su,Bonelli:2010gk}:
  the conformal block is parameterized by $n + (2g - 2 + n)$ parameters,
  where the first factor is from the external momenta and the second from the internal ones.
  In general the generalized matrix model corresponding to this Riemann surface has $n$ $m_{k}$ parameters
  and $g$ parameters including in the term involving the integrals of the holomorphic differentials.
  Since critical points of the potential lead to $(2g-2+n)-1$ filling fraction 
  (where $-1$ comes from the momentum conservation),
  we have the same number of the parameters as the conformal block.

\section{Large $N$ limit}
\label{sec:limit}
  Let us start an analysis of the matrix models introduced in section \ref{sec:DF},
  focusing on the relation with four-dimensional gauge theory.
  One way to study a hermitian matrix model is 
  to make use of the loop equation \cite{David:1990ge,Ambjorn:1990ji,Itoyama:1990mf},
  and take the limit where the size of matrix, $N$, is large.
  By this we can calculate the partition function of the matrix model
  in the iterative way 
  as in \cite{Ambjorn:1992gw,Akemann:1996zr} (see {\it e.g.}~\cite{Marino:2004eq} for a review).
  The systematic study of this method, so-called topological recursion has been performed 
  in \cite{Alexandrov:2003pj,Eynard:2004mh,Eynard:2007kz}, 
  and in \cite{Eynard:2008mz,Chekhov:2009mm,Chekhov:2010xj} for the $\beta$-deformed case.
  An advantage of considering the large $N$ limit (while $g_{s} N$ kept fixed) of the matrix model introduced above 
  is that the limit nicely corresponds to the one where $\epsilon_{1}$ and $\epsilon_{2}$ go to zero 
  in the four-dimensional side, as can be seen from \eqref{epsilonrelation}.
  Thus, this section is devoted to study this limit 
  and see the correspondence with the four-dimensional gauge theory.
    
  In section \ref{subsec:loop}, we derive the loop equation of the $\beta$-deformed matrix model.
  We see that this equation can be interpreted as the Virasoro constraints in the conformal field theory.
  Then we show in section \ref{subsec:spectral} that in the large $N$ limit the spectral curve 
  obtained from the loop equation can be identified with the Seiberg-Witten curve of the corresponding gauge theory.
  The free energy of the matrix model can also be computed and agrees with the prepotential of the gauge theory.
  In section \ref{subsec:looptorus}, we turn to the generalized matrix model on torus, and consider the large $N$ limit.

\subsection{Loop equation}
\label{subsec:loop}
  Let us define the generator of the multi-trace operators as
    \bea
    R(z_{1}, \ldots, z_{k})
     =     (bg_{s})^{k} \sum_{I_{1}} \frac{1}{z_{1} - \lambda_{I_{1}}} \cdots \sum_{I_{k}} \frac{1}{z_{k} - \lambda_{I_{k}}}.
           \label{resolvent}
    \eea
  When $k=1$ this is simply the generator of the single trace operators.
  First of all, we consider the Schwinger-Dyson equation associated to the transformation 
  $\delta \lambda_{I} = \frac{1}{z - \lambda_{I}}$,
  keeping the potential arbitrary
    \bea
    0
    &=&    \frac{1}{Z}
           \int \prod_{I=1}^N d \lambda_I \sum_K \frac{\partial}{\partial \lambda_K}
           \left[ \frac{1}{z - \lambda_K} \prod_{I < J} (\lambda_I - \lambda_J)^{- 2b^2} 
           e^{-\frac{b}{g_s} \sum_I W(\lambda_I)} \right]
           \nonumber \\
    &=&  - \frac{1}{g_s^2} \langle R(z,z) \rangle - \frac{b + \frac{1}{b}}{g_s} \langle R(z)' \rangle
         - \frac{1}{g_s^2} W'(z) \langle R(z) \rangle + \frac{f(z)}{4g_s^2},
           \label{SD}
    \eea
  where $R'$ is the $z$-derivative of the resolvent and 
  we have defined 
    \bea
    f(z)
     =     4 b g_s \left< \sum_I \frac{W'(z)- W'(\lambda_I)}{z - \lambda_I} \right>.
    \eea
  The expectation value is defined as the matrix model average \eqref{average}.
  By multiplying  (\ref{SD}) by $-g_s^2$, we obtain
    \bea
    0
     =     \langle R(z,z) \rangle + (\epsilon_1 + \epsilon_2) \langle R(z)' \rangle
         + W'(z) \langle R(z) \rangle- \frac{f(z)}{4}.
           \label{loopequation}
    \eea
  In the case of the hermitian matrix model $b=i$, the second term vanishes and 
  the equation reduces to the well-known one.

  We now see that this loop equation is interpreted 
  as the Virasoro constraints in the CFT language.
  To see this, let us write the energy-momentum tensor by using the expression \eqref{collective}
    \bea
    g_{s}^{2} T(z)
    &=&  - \left( \frac{1}{4} W'(z)^{2} + \frac{Q}{2} W''(z) + \frac{f(z)}{4} 
         + ({\rm r.h.s.~of}~\eqref{loopequation}) \right).
           \label{T}
    \eea
  The singular part in $z$ only comes from the last term. 
  (We here assume that the potential is a polynomial.)
  Therefore the Virasoro constraint $0 = g_{s}^{2}\langle T(z)|_{{\rm sing}} \rangle$ is equivalent 
  to the loop equation.
  The expectation value of $g_{s}^{2}T(z)$ is simply the first three terms in \eqref{T}.
  
  We now define the ``quantum'' spectral curve as
    \bea
    0
     =     \hat{x}^{2}
         + g_{s}^{2} \langle T(z) \rangle
     =     \langle (\hat{x} + \frac{g_{s}}{\sqrt{2}} \partial \phi)(\hat{x} - \frac{g_{s}}{\sqrt{2}} \partial \phi) \rangle,
           \label{quantumspectral}
    \eea
  where we introduce the commutation relation $[\hat{x}, z] = - Qg_{s}$.

\subsection{Large $N$ limit and Seiberg-Witten theory}
\label{subsec:spectral}
  We now take the large $N$ limit 
  while the filling fractions $\nu_i \equiv b g_s N_i$ ($i=1,\ldots,n-2$) are fixed.
  As we saw in section \ref{subsec:DF}, there are $n-3$ independent filling fractions
  because of the momentum conservation.
  Since both $bg_{s}$ and $g_{s}/b$ send to zero, this limit corresponds to $\epsilon_{1,2} \rightarrow 0$ 
  in the four-dimensional side. 
  
  In this limit the resolvent $\langle R(z,z) \rangle$ is factorized to $\langle R(z) \rangle^{2}$ in the large $N$.
  Therefore the loop equation is written as
    \bea
    0
     =     \langle R (z) \rangle^2 + \langle R (z) \rangle W'(z) - \frac{f(z)}{4},
    \eea
  which is solved as
    \bea
    \langle R(z) \rangle
     =   - \frac{1}{2} \left( W'(z) - \sqrt{W'(z)^{2}+f(z)} \right).
           \label{RN}
    \eea
  The sign has been chosen such that the large $z$ asymptotics agrees with the definition of $R(z)$.
  The spectral curve \eqref{quantumspectral} now becomes ``classical'' because $[ z, x]=0$:
    \bea
    x^{2}
     =     \frac{1}{4} (W'(z)^{2}+f(z)).
           \label{spectral}
    \eea
  It is easy to see that $x = \pm (W'/2 + \langle R \rangle)$ from \eqref{RN} and \eqref{spectral}, 
  which is indeed the classical value of $\hat{x}$ by using \eqref{collective}.  
  Note that the $b$-dependence has disappeared by defining the resolvent as in \eqref{resolvent}.
  Thus, in the large $N$ limit we get the same spectral curve for arbitrary $b$.

  Let us then analyze $f(z)$ by specifying the potential to (\ref{potentialC0n}).
  In this case, 
    \bea
    f(z)
     =     \sum_{k=0}^{n-2} \frac{c_k}{z - w_k},
           \label{f}
    \eea
  where for $k \geq 2$
    \bea
    c_k
     =   - 4 b g_s \left< \sum_I \frac{2 m_k}{\lambda_I - w_k} \right>
     =   - 4 g_s^2 \frac{\partial \log Z}{\partial w_k}
     =   - 4 \frac{\partial F_m}{\partial w_k}.
           \label{ck}
    \eea
  The remaining $c_0$ and $c_1$ can be written in terms of $c_k$ with $k \geq 2$ as follows.
  First of all, due to the equations of motion: $\left< \sum_I W'(\lambda_I) \right> = 0$, 
  the sum of $c_k$ is constrained to vanish $\sum_{k=0}^{n-2} c_k = 0$.
  In order to find another constraint, we consider the asymptotic at large $z$ of the loop equation.
  The asymptotic of the resolvent is $\langle R(z) \rangle  \sim \frac{b g_s N}{z}$, 
  so that the leading terms at large $z$ in the loop equations satisfy
    \bea
    (b g_s N)^2 - (\epsilon_1 + \epsilon_2) b g_s N + b g_s N \sum_{k=0}^{n-2} 2 m_k
    - \sum_{k=0}^{n-2} \frac{w_k c_k}{4}
     =     0.
    \eea
  The leading term of order $1/z$ in $f(z)$ vanishes via the first constraint.
  Thus, we obtain
    \bea
    \sum_{k=0}^{n-2} w_k c_k
     =   - 4\left( \sum_{k=0}^{n-2} m_k + m_{n-1} - g_s Q \right)
           \left( \sum_{k=0}^{n-2} m_k - m_{n-1} \right)
     =:
           M^2,
           \label{crelation2}
    \eea
  where we have used the momentum conservation \eqref{momentumconservationC0n}.
  Therefore, $c_0$ and $c_1$ can be written in terms of $c_k$ (\ref{ck}).
  This means that we have $n-3$ undetermined parameters in the matrix model.

  By substituting the potential the curve \eqref{spectral} is of the form
    \bea
    x^{2}
     =     \sum_{k=0}^{n-2} \frac{m_{k}^{2}}{(z - w_{k})^{2}} + f(z)/4
     =     \frac{P_{2n-4}(z)}{\prod_{k=0}^{n-2}(z - w_{k})^{2}},
    \eea
  where $P_{2n-4}$ is a polynomial of degree $2n-4$, and the residues of $f(z)$ at $z=w_{k}$ \eqref{f} 
  are nontrivial functions of the vacuum values of single trace operators.
  The zeros of $P_{2n-4}$ are the branch points on the $z$-plane, and there are $n-2$ branch cuts.
  Let us define the meromorphic differential $\lambda_m = \frac{x dz}{2 \pi i}$.
  This has simple poles at $z = w_{k}, \infty$ with the residues $m_{k}, m_{n-1}$,
  by observing $\langle R \rangle \sim \frac{g_s N}{z}$ and $W'(z) \sim \sum_{k=0}^{n-2} 2m_{k}/z$
  at large $z$ and by using the momentum conservation.
  By definition, the filling fractions are obtained by the contour integrals of this differential
    \bea
    \nu_{i}
     =     \oint_{C_{i}} dz \lambda_{m}.
           \label{contour}
    \eea 
  where $C_{i}$ ($i=1, \ldots, n-2$) are the contours around the branch cuts.
  These equations relate the vevs of the single trace operators included in $f(z)$ with the filling fraction $\nu_{i}$.
  
  This is exactly the form of the Seiberg-Witten curve of the $SU(2)$ linear quiver gauge theory, 
  $x^{2} = \phi_{2}$ where $\phi_{2}$ is a quadratic meromorphic differential on a sphere.
  Moreover the differential defined above is identified with the Seiberg-Witten differential 
  $\lambda_{{\rm SW}} = \frac{\sqrt{\phi_{2}}dz}{2 \pi i}$.
  Indeed, as proposed in section \ref{sec:DF}, the filling fractions are mapped to the vacuum expectation values 
  of the vector multiplet scalars, since
  in the Seiberg-Witten theory these are given by contour integrals of the Seiberg-Witten differential
  exactly in the same way as \eqref{contour}.
  For the case with $n=4$ associated with the $SU(2)$ gauge theory with four fundamental hypermultiplets,
  the precise identification between the vevs of single trace operators
  and the Coulomb moduli parameter has been worked out in \cite{Eguchi:2009gf}.
  In \cite{Schiappa:2009cc}, the standard saddle point analysis developed in \cite{Ambjorn:1992gw} 
  has been applied to determine the spectral curve, in particular the positions of branch cuts.
  
  This is in agreement with the argument in \cite{Alday:2009aq} that 
  the $\phi_{2}$ appearing in the Seiberg-Witten curve can be identified with the vacuum expectation value of the 
  energy-momentum tensor of the Virasoro CFT
    \bea
    \phi_{2}(z)
     =     g_{s}^{2} \langle T(z) \rangle|_{\epsilon_{1,2} \rightarrow 0},
    \eea
  by recalling our definition of the spectral curve \eqref{quantumspectral}.

\paragraph{Free energy}
  So far we have seen the identification of the spectral curve of the matrix model and the Seiberg-Witten curve
  of the gauge theory.
  However, it is still not straightforward to see the equivalence of the free energy of the former 
  with the prepotential of the latter,
  because the special geometry relation of the Seiberg-Witten theory: 
  $a=\oint_{A}\lambda_{{\rm SW}}$ and $\frac{\partial \CF}{\partial a} = \oint_{B}\lambda_{{\rm SW}}$,
  where $\CF$ is the prepotential, is not manifest in the matrix model.
  The saddle point analysis of the matrix model can be used to obtain the free energy
  and the equation like the (second) spacial geometry relation, as in \cite{Eguchi:2009gf}.
  However here let us shortly see a more direct approach to the free energy
  for the $n=4$ case considered in \cite{Eguchi:2010rf}.
  
  Recall the relation \eqref{ck}.
  In the $n=4$ case with $w_{2}=q$, this is
    \bea
    \frac{\partial F_m}{\partial q}
     =   - \frac{c_2}{4}
           \label{freeenergyderivative4}
    \eea
  Therefore, what we need to do is to calculate $c_{2}$.
  (Actually we can only derive the $q$ dependent part of the free energy by this method.) 
  As we discussed in the previous subsection, the parameters $c_{0}$, $c_{1}$ and $c_{2}$ in $f(z)$ are related 
  by $\sum c_{i} = 0$ and \eqref{crelation2} $c_1 + qc_2 =  4m_3^2-4(\sum_{i=0}^2 m_i)^2$ 
  (when $\epsilon_{1,2} = 0$).
  Thus, we have $(1-q)c_{2} = 4(\sum_{i=0}^2 m_i)^2 - 4m_3^2  - c_0$.
  Below we will compute $c_{0}$ by writing the spectral curve in terms of it.
  
  In what follows, we consider the simple case where all the hypermultiplet masses are equal to $m$:
  i.e. $m_0 = m_3 = 0$ and $m_{1}=m_{2}=m$.
  In this case, the polynomial $P_{4}$ in the spectral curve is reduced to degree $3$: 
  $P_3(z) = C z (z - z_+)(z - z_-)$, 
  where we have introduced $C = c_0 q/4$ and 
    \bea
    z_\pm
     =     \frac{1}{2} \left( 1+q - (1-q)^2 \frac{m^2}{C} 
         \pm (1-q) \sqrt{1 - 2(1+q) \frac{m^2}{C} + (1-q)^2 \frac{m^4}{C^2} } \right).
    \eea
  By taking the $C$ derivative of $xdz$, we get the holomorphic differential with 
    \bea
    \frac{\partial}{\partial C} xdz
     =     \frac{1}{2 \sqrt{C z_+}} \frac{dz}{\sqrt{z (1-z)(1 - k^2 z)}}, ~~~
    k^2
     =     \frac{z_-^2}{q}.
    \eea
  Since the contour integral of this differential gives the $C$ derivative of the filling fraction $\nu_{1}$
  which has been identified with the vevs $a$ by $a = bg_s N_1$.
  Thus by expanding in $\frac{m^2}{C}$ and integrating over $C$, we obtain
    \bea
    a
     =     \sqrt{C} \left( h_0(q) - h_1(q) \frac{m^2}{C} - \frac{h_2(q)}{3} \frac{m^4}{C^2} 
         + \CO \left( \frac{m^6}{C^3} \right) \right),
    \eea
  where $h_i(q)$ depend only on $q$ and are given in \cite{Eguchi:2010rf}.
  By solving for $C$, substituting it into (\ref{freeenergyderivative4}), and integrating over $q$, 
  we finally obtain the free energy
    \bea
    F_m
    &=&    (a^2 - m^2) \log q + \frac{a^4+6 a^2 m^2 + m^4}{2 a^2} q
           \nonumber \\
    & &  + \frac{13 a^8 + 100 m^2 a^6 + 22 m^4 a^4 - 12 m^6 a^2 + 5 m^8}{64a^6}q^2
         + \CO(q^3).
    \eea
  This agrees with the prepotential of the $SU(2)$ gauge theory with four fundamental hypermultiplets.
  The latter can be obtained from the Nekrasov partition function of $U(2)$ gauge theory 
  by subtracting the terms coming from the $U(1)$ factor.
  
\paragraph{Subleading order of large $N$ expansion}
  It is interesting to check the subleading order in the large $N$ (small $\epsilon_{1}$, $\epsilon_{2}$) expansion.
  On the four-dimensional side, the Nekrasov partition function is expanded as
    \bea
    \CF
     :=     \epsilon_{1} \epsilon_{2} \ln Z_{{\rm Nek}}
     =     \CF_{0} + (\epsilon_{1} + \epsilon_{2}) H + \epsilon_{1} \epsilon_{2} F_{1}
         + (\epsilon_{1}+\epsilon_{2})^{2} G + \ldots 
    \eea
  Subleading terms $H$, $F_{1}$ and $G$ can be obtained from the geometric data of the Seiberg-Witten theory.
  (See \cite{Nakajima:2003uh} for detail.)
  The matrix model analysis for the subleading orders can also be done.
  In particular, it was shown that the corresponding parts of the free energy agrees 
  with $F_{1}$ in \cite{Schiappa:2009cc} and with $H$ and $G$ in \cite{Itoyama:2011mr,Nishinaka:2011aa}.
  For generic $b$, this expansion of the matrix model was compared \cite{Morozov:2010cq}
  with the finite $N$ calculation which will be explained in section \ref{sec:finiteN}.
  
  The method using the topological recursion \cite{Alexandrov:2003pj,Eynard:2004mh,Eynard:2007kz} would be useful. 
  In particular, the calculation of the partition function of the $\beta$-deformed matrix model
  with the logarithmic potential was considered in \cite{Chekhov:2010xj,Brini:2010fc} in this context.

\subsection{Higher genus case}
\label{subsec:looptorus}
  Let us turn to the generalized matrix model corresponding to the torus \eqref{holo},
  and derive the loop equation.
  We then see the equivalence of the spectral curve obtained by taking the large $N$ limit 
  and the Seiberg-Witten curve \cite{Bonelli:2011na}.
  
  We now define the toric version of the resolvent
    \bea
    R(z_1, \ldots, z_k)
     =     (b g_s)^k \sum_{I_1} \frac{\theta_1'(z_1 - \lambda_{I_1})}{\theta_1(z_1 - \lambda_{I_1})} \ldots 
           \sum_{I_k} \frac{\theta_1'(z_k - \lambda_{I_k})}{\theta_1(z_k - \lambda_{I_k})}.
           \label{resolventtorusn}
    \eea
  From the Schwinger-Dyson equation for an arbitrary transformation 
  $\delta \lambda_K = \frac{\theta_1'(z - \lambda_K)}{\theta_1(z - \lambda_K)}$, we derive
    \bea
    0
    &=&    g_s^2 \left< \sum_I \left( \frac{\theta_1'(z - \lambda_I)}{\theta_1(z - \lambda_I)} \right)^2 \right>
         - g_s^2 \left< \sum_I \frac{\theta_1^{''}(z - \lambda_I)}{\theta_1(z - \lambda_I)} \right>
         - b g_s W'(z) \left< \sum_I \frac{\theta_1'(z - \lambda_I)}{\theta_1(z - \lambda_I)} \right>
           \nonumber \\
    & &  + t(z)
         - 2 b^2 g_s^2 \left< \sum_{I<J} \frac{\theta_1'(\lambda_I - \lambda_J)}{\theta_1(\lambda_I - \lambda_J)}
           \left( \frac{\theta_1'(z - \lambda_I)}{\theta_1(z - \lambda_I)}
         - \frac{\theta_1'(z - \lambda_J)}{\theta_1(z - \lambda_J)} \right) \right>,
           \label{SD1}
    \eea
  where we have multiplied by $g_s^2$ and defined
    \bea
    t(z)
     =     b g_s \left< \sum_I \frac{\theta_1'(z - \lambda_I)}{\theta_1(z - \lambda_I)} (W'(z)-W'(\lambda_I)) \right>.
    \eea
  By using the formula of the theta function and, after some algebra, we obtain \cite{Bonelli:2011na}
    \bea
    0
    &=&  - \left< R(z,z) \right> - (\epsilon_{1}+\epsilon_{2}) \left< R'(z) \right>
         - W'(z) \left< R(z) \right>
         + b^2 g_s^2 N \left< \sum_I \frac{\theta_1^{''}(z - \lambda_I)}{\theta_1(z - \lambda_I)} \right>
           \nonumber \\
    & &  + t(z) + b^2 g_s^2 \left< \sum_{I<J} \frac{\theta_1^{''}(\lambda_I - \lambda_J)}{\theta_1(\lambda_I - \lambda_J)} \right>
         + 3 b^2 g_s^2 \eta_1 N(N-1),
           \label{SD2}
    \eea
  where $\eta_1 = 4 \frac{\partial \ln \eta}{\partial \ln q}$.
  This equation is valid for an arbitrary potential.
  
  Let us now focus on the potential (\ref{c_action}).
  By rewriting $t(z)$ we finally obtain 
    \bea
    0
    &=&  - \left< R(z,z) \right> - (\epsilon_{1}+\epsilon_{2}) \left< R'(z) \right>
         + W'(z) \left< R(z) \right> - 3 b g_s (N+1) \eta_1 \sum_k m_k
           \label{SD4} \\
    & &  - 2 b g_s \sum_{k=1}^n m_{k} \frac{\theta_1'(z - w_k)}{\theta_1(z - w_k)} 
           \left< \sum_{I} \frac{\theta_1'(\lambda_{I} - w_k)}{\theta_1(\lambda_{I} - w_k)} \right>
         + b g_s N \sum_k m_k \frac{\theta_1^{''}(z - w_k)}{\theta_1(z - w_k)}
         + 4 g_s^2 \frac{\partial \ln Z}{\partial \ln q}.
           \nonumber
    \eea
  
  Let us now see the spectral curve in the large $N$ limit.
  For simplicity we consider the $n=1$ case, and take $w_1 = 0$.
  In this case, it is easy to see that the loop equation reduces to
    \bea
    0
    &=&  - x^2 + m_1^{2} \CP(z) - 4 u,
           \label{SD7}
    \eea
  where $\CP$ is the Weierstrass function, $x= \pm (\langle R \rangle + W'/2)$, and
    \bea
    u
     =   - \pi^2 a^2
         + \frac{\partial}{\partial \ln q} \left( F_{0}
         - m_1^{2} \ln \eta \right).
           \label{utorusn}
    \eea
  We defined the free energy as 
  $F_{0} = \lim_{\epsilon_{1,2} \rightarrow 0} (\epsilon_{1}\epsilon_{2}) \ln Z$.
  This is indeed the Seiberg-Witten curve of the $SU(2)$ $\CN=2^{*}$ theory.

\section{Nekrasov-Shatashvili limit}
\label{sec:NS}
  It has long been known that the Seiberg-Witten theory is related with the classical integrable system 
  \cite{Gorsky:1995zq,Martinec:1995by,Nakatsu:1995bz,Donagi:1995cf,Itoyama:1995nv,Itoyama:1995uj}.
  The connection has been considered in \cite{Gaiotto:2009hg}
  from the recent perspective of the 6d (2,0) theory compactification. 
  A review can be found in \cite{N}.
  Quite remarkably it was proposed in \cite{Nekrasov:2009rc} that the gauge theory on the $\Omega$ background with
  $\epsilon_2 \rightarrow 0$ while $\epsilon_1$ kept fixed is related with the quantization of the integrable system.
  In this section, we consider this limit from the matrix model side.
  The limit is translated to $b \rightarrow \infty$ and $g_s \rightarrow 0$ 
  with $b g_s$, $g_{s} \alpha_{k}$ and $g_{s} N_{i}$ kept finite,
  and corresponds to the semiclassical limit in the CFT.
  
  In this limit, the leading order part of the free energy is obtained from the value of the critical points 
  which solve the equations of motion \eqref{eom}, as in the large $N$ limit.
  We note that two terms in \eqref{eom} are of the same order in the limit 
  because $N$ and $\epsilon_1$ are kept finite.
  Let us then consider the loop equation (\ref{loopequation}).
  Again, in this limit, 
  the connected part of (\ref{resolvent}) can be ignored: 
  $\langle R(z,z) \rangle \rightarrow \langle R(z) \rangle^2$.
  Taking this into account, (\ref{loopequation}) becomes
    \bea
    0
     =     \langle \tilde{R}(z) \rangle^2 + \epsilon_1 \langle \tilde{R}(z)' \rangle
         + \langle \tilde{R}(z) \rangle W'(z) - \frac{\tilde{f}(z)}{4},
    \eea
  where $\tilde{R}$ and $\tilde{f}$ are $R|_{\epsilon_2 \rightarrow 0}$ and $f|_{\epsilon_2 \rightarrow 0}$ respectively.
  In the following, we will omit the tildes of $R$ and $f$.
  Then, in terms of $x = \langle R(z) \rangle + W'(z)/2$, the equation becomes
  \cite{Eynard:2008mz,Chekhov:2009mm,Mironov:2009ib}
    \bea
    0
     =   - x^2 - \epsilon_1 x' + U(z),
    \eea
  where
    \bea
    U(z)
     =     \frac{1}{4} \left (W'(z)^2 + 2\epsilon_1 W^{''}(z) + f(z) \right).
           \label{U}
    \eea
  This is a Ricatti type equation.
  It is then easy to see that this can be written as the Schr\"odinger-type equation:
    \bea
    0
     =   - \epsilon_1^2 \frac{\partial^2}{\partial z^2} \Psi(z) + U(z) \Psi(z),
           \label{diffC0n}
    \eea
  where the ``wave function'' $\Psi (z)$ is defined by
    \bea
    \Psi(z)
     =     \exp \left( \frac{1}{\epsilon_1} \int^{z} x(z') dz' \right).
    \eea
  This indicates the relation between the $\beta$-deformed matrix model and quantum integrable system. 
  
  Note that the quantum spectral curve indeed leads to the same conclusion.
  Eq.~\eqref{quantumspectral} becomes in this limit
    \bea
    0
     =     \hat{x}^{2} - U(z).
    \eea
  These variables are not commutative $[\hat{x}, z] = - \epsilon_{1}$. 
  Thus $\hat{x}=-\epsilon_{1}\frac{\partial}{\partial z}$ which leads to \eqref{diffC0n}.
  
  In \cite{Teschner:2010je, Maruyoshi:2010iu}, it was shown that the the conformal block on a sphere
  with the additional insertion of the degenerate fields $V_{-\frac{1}{2b}}(z) = e^{- \frac{\phi(z)}{\sqrt{2}b}}$ 
  captures the quantization of the integrable systems.
  The details can be found in \cite{T}.
  (The similar relation between the affine $SL(2)$ conformal block and integrable system has been found 
  in \cite{Alday:2010vg}.)
  This has an interpretation in the 4d gauge theory as an insertion of a surface operator \cite{Alday:2009fs}
  (see \cite{Gu} for a review on this part.)
  Similar connections with the integrable system has been considered 
  in \cite{Dorey:2011pa,Chen:2011sj,Bulycheva:2012ct,Chen:2013jtk}.
  In the following, we will show that 
  under the identification of the $\beta$-deformed matrix model $Z$ with the $n$-point conformal block, 
  the integral representation of the conformal block with degenerate field insertions can be written 
  in terms of the resolvent of the original matrix model \cite{Marshakov:2010fx,Bonelli:2011na}, 
  in the $\epsilon_2 \rightarrow 0$ limit.
  We note that the similar analysis was done in \cite{Aganagic:2011mi} from the topological string viewpoint.

  Let us consider the integral representation of the ($n+\ell$)-point conformal block 
  where $\ell$ degenerate fields are inserted
    \bea
    Z_{\ell}
    &=&    \left< \prod_{i=1}^\ell V_{\frac{1}{2b}}(z_i) \left( \int d \lambda e^{\sqrt{2}b \phi(\lambda)} \right)^N ~
           \prod_{k=0}^{n-1} V_{\frac{m_k}{g_s}}(w_k) \right>
           \nonumber \\
    &=&    \prod_{i<j} (z_i - z_j)^{- \frac{1}{2 b^2}} 
           \prod_{0 \leq k < \ell \leq n-2} (w_k - w_\ell)^{- \frac{2 m_k m_\ell}{g_s^2}} 
           \prod_{i=1}^\ell \prod_{k=0}^{n-2} (z_i - w_k)^{\frac{m_k}{b g_s}}
           \nonumber \\
    & &    ~~~~~~\times
           \int \prod_{I=1}^N d \lambda_I \prod_{I<J} (\lambda_I - \lambda_J)^{- 2b^2}
           \prod_{I} \prod_{k=0}^{n-2} (\lambda_I - w_k)^{-\frac{2 b m_k}{g_s}} \prod_{i=1}^\ell (z_i - \lambda_I),
           \label{degenerateC0n}
    \eea
  where we have taken $w_{n-1}$ to infinity and omitted the factor including this, as we have done above.
  The momentum conservation is however modified by the degenerate field insertion as
    \bea
    \sum_{k=0}^{n-1} m_k - \frac{\ell g_s}{2 b} + b g_s N
     =     g_s Q.
           \label{momentumC0n+ell}
    \eea
  By dividing by $Z$ and taking a log, we obtain
    \bea
    \log \frac{Z_{\ell}}{Z}
     =   - \frac{1}{2 b^2} \sum_{i<j} \log (z_i - z_j) 
         + \sum_i \frac{W(z_i)}{2 b g_s} + \log \left< \prod_{i, I} (z_i - \lambda_I) \right>,
           \label{logratio}
    \eea
  where the potential $W(z)$ is the same as (\ref{potentialC0n}).
  Notice that the expectation value is defined with the modified momentum conservation (\ref{momentumC0n+ell}).
  By defining $e^L = \prod_{i, I} (z_i - \lambda_I)$, 
  we notice that
  $L = \sum_{i, I} \log (z_i - \lambda_I) = \sum_{i, I} \int^{z_i} \frac{d z_i'}{z_i' - \lambda_I}$,
  where we have ignored irrelevant terms due to the end points of the integrations. 
  Then, we use that the expectation value of $e^L$ can be written as 
  $\log \left< e^L \right> = \sum_{k=1}^\infty \frac{1}{k!} \left< L^k \right>_{conn}$ \cite{Marshakov:2010fx}, 
  where $\left< \ldots \right>_{conn}$ means the connected part of the correlator,  
  $\langle L^2 \rangle_{conn} = \langle L^2 \rangle - \langle L \rangle^2$, etc, 
  while $\langle L \rangle_{conn} = \langle L \rangle$.
  Thus, the last term in the right hand side of (\ref{logratio}) can be expressed as
  $\sum_{k=1}^\infty \frac{1}{k!} 
  \left< \left( \sum_{i,I} \int^{z_i} \frac{dz'}{z' - \lambda_I} \right)^k \right>_{conn}$.

  In the limit where $\epsilon_2 \rightarrow 0$, 
  the terms with $k>1$ of the previous expression are subleading contributions 
  compared with the $k=1$ terms 
  since the connected part of the expectation value can be ignored.
  Also the first term in the right hand side of (\ref{logratio}) can be neglected in this limit.
  Thus, we obtain
    \bea
    \frac{Z_{\ell}}{Z}
     \rightarrow     \prod_{i=1}^\ell \Psi_i(z_i), ~~~
    \Psi_i (z_i)
     =     \exp \left( \frac{1}{\epsilon_1} \int^{z_i} x(z') dz' \right).
           \label{PsiZ5}
    \eea
  This indicates that the properties of the conformal block with degenerate field insertions
  are build in the resolvent of the matrix model in the $\epsilon_2 \rightarrow 0$ limit.
  This property of ``separation of variables" agrees with the corresponding result of the Virasoro conformal block 
  as in \cite{Kozcaz:2010af, Teschner:2010je}.
  Furthermore, this $\Psi$ with $\ell=1$ is exactly the one which satisfied the Schr\"odinger equation \eqref{diffC0n}.
  
  In summary, we have seen that the integral representation corresponding to the insertion of the degenerate fields
  into the Virasoro conformal block satisfies the Schr\"odinger equation,
  whose potential can be obtained from the loop equation.

\paragraph{Relation with Gaudin model}
  The above argument is applicable for an arbitrary potential $W(z)$.
  Here we return to the logarithmic one (\ref{potentialC0n}) 
  and see the relation \cite{Eynard:2008mz,Bonelli:2011na,Bourgine:2012gy,Bourgine:2012bv} with the Gaudin Hamiltonian.
  In this case (\ref{U}) becomes \cite{Bonelli:2011na}
    \bea
    U(z)
     =     \sum_{k=0}^{n-2} \frac{m_k (m_k + \epsilon_1)}{(z - w_k)^2}
         + \sum_k \frac{H_k}{z - w_k}
         - \sum_{k=0}^{n-2} \frac{c_k/4}{z-w_k}
           \label{UC0n}
    \eea
  where
    \bea
    H_k
     =     \sum_{\ell(\neq k)} \frac{2 m_k m_\ell}{w_k - w_\ell}.
    \eea
  $U(z)$ is indeed the vacuum expectation value of Gaudin Hamiltonian.
  In particular, $H_k - c_k/4$ are the vacuum energies of the quantum Hamiltonians.
  
  So far, we discussed the case corresponding to the CFT on the sphere.
  For the toric case, it has been shown that the loop equation of the generalized matrix model 
  in subsection \ref{subsec:looptorus} gives the Hamiltonian of the Hitchin system 
  on the torus in \cite{Bonelli:2011na}.
  In particular the $n=1$ case leads to the elliptic Calogero-Moser model.

\section{Finite $N$ analysis}
\label{sec:finiteN}
  In the previous section we considered the large $N$ limit and the Nekrasov-Shatashvili limit 
  of the $\beta$-deformed matrix model.
  Here we will see a different expansion of the matrix model partition function 
  in the complex structures of the Riemann surface.
  We calculate each order of the expansion by performing the direct integration.
  Indeed, this expansion is more useful to compare 
  with the Virasoro conformal block and the Nekrasov partition function.
  We first review the conformal block of the Virasoro algebra in subsection \ref{subsec:Virasoro}.
  Then we analyze the integral representation in subsection \ref{subsec:finiteN}.

\subsection{Virasoro conformal block}
\label{subsec:Virasoro}
  Let us review the Virasoro algebra and the conformal block \cite{Belavin:1984vu}.
  (See {\it e.g.} \cite{Marshakov:2009gs,Mironov:2009dr} for detailed computations.)
  We consider the conformal symmetry generated by the holomorphic energy-momentum tensor $T(z)$
  with 
    \bea
    T(z) 
     =     \sum_{n=-\infty}^{\infty} \frac{L_{n}}{z^{n+2}}.
    \eea
  The Virasoro algebra is
    \bea
    \left[ L_m, L_n \right]
    &=&    (m - n) L_{m+n} + \frac{c}{12} (m^3 - m) \delta_{m+n, 0},
    \eea
  and we consider the case with Liouville like central charge $c = 1 + 6 Q^2$ where $Q = b + 1/b$.
  
  The primary field $V_\alpha(x)$ corresponds to the highest weight vector satisfying
    \bea
    L_n V_\alpha
     =     0, ~~~
    L_0 V_\alpha
     =     \Delta_{\alpha} V_\alpha,
    \eea
  where $n>0$. 
  The conformal dimension of the primary is $\Delta_{\alpha} = \alpha (Q - \alpha)$.
  By state-operator correspondence we denote the primary state with $\Delta_{\alpha}$ by $| \Delta \rangle$.
  Then the Verma module $\mathcal{V}$ is formed by the descendants $V_{Y,\alpha} = :L_{-Y} V_{\alpha}:$ 
  which are obtained by acting with the raising operators 
  $L_{-Y} = (L_{-y_1})^{n_{1}} (L_{-y_2})^{n_{2}} (L_{-y_3})^{n_{3}} \cdots$,
  where $\{y_i\}$ are positive integers with $y_1 < y_2 < \cdots$.
  Below we use the shorthand notation to denote $Y$:
  $Y = [\cdots y_{3}^{n_{3}}y_{2}^{n_{2}} y_{1}^{n_{1}}]$, {\it e.g.,} for $y_{1}=1$ and $n_{1}=2$, $Y=[1^{2}]$.
  Let us denote the sum of $n_{i}y_{i}$ as $|Y|$.
  The dimension of the descendant is $\Delta_{\alpha} + |Y|$, and we call $|Y|$ as level.
  
  The OPE of these operators is given by 
    \bea
    V_{Y_{1}, \alpha_{1}}(q) V_{Y_{2},\alpha_{2}}(0)
     =     \sum q^{\Delta -\Delta_{1} - \Delta_{2}-|Y_{1}|-|Y_{2}|} C^{\alpha}_{\alpha_{1},\alpha_{2}}
           \sum_{Y} q^{|Y|} \beta^{\Delta, Y}_{\Delta_{1},Y_{1}; \Delta_{2},Y_{2}} V_{Y, \alpha}(0),
    \eea
  where $\Delta$ is the conformal dimension of $V_{\alpha}$.
  $C^{\alpha}_{\alpha_{1},\alpha_{2}}$ depends on the dynamics of a two-dimensional theory
  while $\beta^{\Delta, Y}_{\Delta_{1},Y_{1}; \Delta_{2},Y_{2}}$ is determined from the Virasoro algebra only
  and depends on the conformal dimensions and central charge.
  We will focus on the latter and ignore the factors $C_{\alpha_{1},\alpha_{2}}^{\alpha_{3}}$ coming from the dynamics.
  
  Let us now define the two-point function
    \bea
    Q_{\Delta}(Y_{1},Y_{2})
     =     \langle \Delta | L_{Y_{1}} L_{-Y_{2}} | \Delta \rangle.
    \eea
  This is symmetric under the exchange of $Y_{1}$ and $Y_{2}$,
  and vanishes unless $|Y_{1}| = |Y_{2}|$.
  By using this, $\beta^{\Delta, Y}_{\Delta_{1},Y_{1}; \Delta_{2},Y_{2}}$ can be written 
  in terms of the three-point function $\gamma$:
    \bea
    \gamma_{\Delta_{1},\Delta_{2}, \Delta_{3}}(Y_{1},Y_{2},Y_{3})
    &=&    \langle  V_{Y_{1},\alpha_{1}}(\infty) V_{Y_{2},\alpha_{2}}(1) V_{Y_{3},\alpha_{3}}(0) \rangle
           \nonumber \\
    &=&    \sum_{Y'} \beta^{\Delta_{3}, Y}_{\Delta_{1},Y_{1}; \Delta_{2},Y_{2}} Q_{\Delta_{3}}(Y',Y_{3}).
    \eea
  When $|Y_{1}|=|Y_{2}| = \emptyset$, the expressions for the $\beta$ and $\gamma$ can be simplified.
  Thus we define in particular
    \bea
    \gamma_{\Delta_{1},\Delta_{2}, \Delta_{3}}(Y) 
    &=&    \gamma_{\Delta_{1},\Delta_{2}, \Delta_{3}}(\emptyset,\emptyset,Y),
           \nonumber \\
    \beta^{\Delta_{3}}_{\Delta_{1}, \Delta_{2}} (Y)
    &=&    \beta^{\Delta_{3}, Y}_{\Delta_{1},\emptyset; \Delta_{2},\emptyset}.
    \eea
  
  These $Q$ and $\gamma$ can be computed order by order in the level.
  Let us give a few results of the computation for later convenience:
    \bea
    Q_{\Delta} ([1],[1])
    &=&    2 \Delta, 
           \nonumber \\
    Q_{\Delta} ([2],[2])
    &=&    4 \Delta + c/2, ~~~
    Q_{\Delta} ([2],[1^{2}])
     =     6 \Delta, ~~~
    Q_{\Delta} ([1^{2}],[1^{2}])
     =     4 \Delta(1 + 2 \Delta),
           \nonumber \\
    &\vdots&
           \nonumber \\
    \gamma_{\Delta_{1},\Delta_{2}, \Delta_{3}}([1])
    &=&    \Delta_{1} + \Delta_{3} - \Delta_{2},
           \nonumber \\
    \gamma_{\Delta_{1},\Delta_{2}, \Delta_{3}}([2])
    &=&    2\Delta_{1} + \Delta_{3} - \Delta_{2},~~~
    \gamma_{\Delta_{1},\Delta_{2}, \Delta_{3}}([1^{2}])
     =     (\Delta_{1} + \Delta_{3} - \Delta_{2})(\Delta_{1} + \Delta_{3} - \Delta_{2} +1),
           \nonumber \\
    & \vdots&
    \eea
  We used $[L_{n},V_{\alpha}(z)] = z^{n} (z \partial_{z} + (n+1)\Delta) V_{\alpha}(z)$ 
  following from the conformal Ward identities when computing the three-point function.
  From this we can calculate $\beta$ as
    \bea
    \beta^{\Delta_{3}}_{\Delta_{1}, \Delta_{2}} ([1])
    &=&    \frac{\Delta_{1} + \Delta_{3} - \Delta_{2}}{2 \Delta_{3}}.
           \label{beta1}
    \eea
  
  Now we can write down the conformal block in terms of these functions.
  Let us focus on the four-point conformal block which we refer to as $\CB$.
  By translation symmetry we put three points at $0,1$ and $\infty$.
  Thus the conformal block is written in terms of the cross ratio $q$ 
  which is the position of the remaining vertex operator.
  Then the conformal block has the following structure:
    \bea
    \CB
     =     \sum_{k=0}^{\infty} \CB_{k} q^{k}, ~~~~~
    \CB_{k}
     =     \sum_{|Y|=|Y'|=k} \gamma_{\Delta_{0},\Delta_{2}, \Delta}(Y) Q_{\Delta}^{-1}(Y,Y')
           \gamma_{\Delta_{1},\Delta_{3},\Delta}(Y'),
    \eea
  and $\CB_{0} =1$.
  The conformal block is computed by order by order. 
  E.g., the first order coefficient $\CB_{1}$ is computed as
    \bea
    \CB_1
     =     \frac{(\Delta + \Delta_0 - \Delta_2)(\Delta + \Delta_1 - \Delta_3)}{2 \Delta}.
    \eea

\subsection{Finite $N$ matrix model}
\label{subsec:finiteN}
  Now we consider the integral representation.
  Let us first see that the prescription for the momentum conservation at the vertex \eqref{momcon} is indeed 
  the correct one by checking the equivalence of the three-point functions.
  To see this, we consider the following OPE in the free scalar theory
    \bea
    & &    :L_{-Y_{1}} V_{\alpha_{1}}(q): :L_{-Y_{2}}V_{\alpha_{2}}(0): \prod_{I=1}^{N} \int_{0}^{q} d \lambda_{I} :e^{\sqrt{2} b \phi(\lambda_{I})}:
           \nonumber \\ 
    & &    ~~~~~~~~~~~~~~
     =     C \sum_{Y} q^{|Y|} \beta^{\Delta_{\alpha_{1} + \alpha_{2}+bN}, Y}_{\Delta_{1}, Y_{1};\Delta_{2},Y_{2}}\Big|_{{\rm free}}
           :L_{-Y} V_{\alpha_{1} + \alpha_{2}+bN}(0):,
           \label{3pointfree}
    \eea
  where $C$ is an irrelevant factor normalizing 
  $\beta_{\Delta_{1},\emptyset; \Delta_{2} \emptyset}^{\Delta_{\alpha_{1}+\alpha_{2}+bN},\emptyset}|_{{\rm free}} = 1$.
  The coefficient $\beta|_{{\rm free}}$ corresponds to the three-point function.
  Thus, it is natural to propose that \cite{Mironov:2010ym}
    \bea
    \beta^{\Delta, Y}_{\Delta_{1},Y_{1}; \Delta_{2},Y_{2}}
     =     \beta^{\Delta_{\alpha_{1} + \alpha_{2}+bN}, Y}_{\Delta_{1},Y_{1}; \Delta_{2},Y_{2}} \Big|_{{\rm free}},
           \label{propmmm}
    \eea
  under the identification of the internal momenta $\alpha = \alpha_{1}+\alpha_{2}+bN$,
  where the left hand side is the one obtained in the previous subsection.

  Let us focus on the case with $Y_{1}=Y_{2}=\emptyset$ and analyze the right hand side 
  of \eqref{3pointfree} further.
  By calculating the OPE in the free field theory, we obtain
    \bea
    & &    V_{\alpha_{1}}(q) V_{\alpha_{2}}(0) \prod_{I=1}^{N} \int_{0}^{q} d \lambda_{I} 
           :e^{\sqrt{2} b \phi(\lambda_{I})}: 
           \label{3pointfree2} \\  
    & &    
     =     q^{-2\alpha_{1}\alpha_{2}} \prod_{I=1}^{N} \int_{0}^{q} d \lambda_{I}
           \prod_{I<J} (\lambda_{I} - \lambda_{J})^{-2b^{2}} \prod_{I=1}^{N} \lambda_{I}^{-2b\alpha_{2}} 
           (q - \lambda_{I})^{-2b\alpha_{1}} :e^{\sqrt{2}(\alpha_{1}\phi(q) +\alpha_{2}\phi(0) + b\sum_{I} \phi(\lambda_{I}))}:
           \nonumber \\
    & &    
     =     q^{\sigma} \prod_{I=1}^{N} \int_{0}^{1} d x_{I} 
           \prod_{I<J} (x_{I} - x_{J})^{-2b^{2}} \prod_{I=1}^{N} x_{I}^{-2b\alpha_{2}} (1 - x_{I})^{-2b\alpha_{1}} 
           \sum_{Y,Y'} q^{|Y|-|Y'|} H_{Y,Y'} x^{Y'} :V_{Y,\alpha}(0):.
           \nonumber 
    \eea
  In the last equality we have changed the variables $\lambda_{I} = q x_{I}$ and defined $H_{Y,Y'}$ such that 
    \bea
    :e^{\sqrt{2}(\alpha_{1}\phi(q) +\alpha_{2}\phi(0) + b\sum_{I} \phi(\lambda_{I}))}:
     =     \sum_{Y,Y'} q^{|Y|-|Y'|} H_{Y,Y'} \lambda^{Y'} :L_{-Y} e^{\sqrt{2}(\alpha_{1}+\alpha_{2}+bN)\phi(0)}:,
           \label{H}
    \eea
  where $\lambda^{Y} = \prod_{I} \lambda_{I}^{y_{I}}$ 
  for the partition $Y=[y_{N},\ldots,y_{1}]$ with $y_{1} \leq y_{2} \leq \ldots$.
  We sum over all the possible $Y$ and $Y'$ with $|Y| \geq |Y'|$.
  By defining the following multiple integral
    \bea
    \langled x^{Y} \rangled_{N}
     =     \prod_{I=1}^{N} \int_{0}^{1} d x_{I} 
           \prod_{I<J} (x_{I} - x_{J})^{-2b^{2}} x^{Y} \prod_{I=1}^{N} x_{I}^{-2b\alpha_{2}} (1 - x_{I})^{-2b\alpha_{1}}, 
           \label{Sel} 
    \eea
  the three-point function $\beta$ from the free scalar field theory is thus
    \bea
    \beta^{\Delta_{\alpha_{1}+\alpha_{2}+bN}}_{\Delta_{1}, \Delta_{2}}(Y) \Big|_{{\rm free}}
     =     \sum_{Y',|Y'| \leq |Y|} H_{Y,Y'} \frac{\langled x^{Y'} \rangled_{N}}{\langled 1\rangled_{N}}.
    \eea
  The multiple integral \eqref{Sel} is of the Selberg type. 
  We will give results of the integration in appendix \ref{sec:Selberg}.
  Thus, the right hand side is in principle calculable.
  
  Let us check the equivalence of the first order.
  In this case $Y=[1]$, $H_{[1],\emptyset} = \frac{\alpha_{1}}{\alpha_{1}+\alpha_{2} + bN}$
  and $H_{[1],[1]} = \frac{bN}{\alpha_{1}+\alpha_{2} + bN}$.
  Combining the formula for $\langled x^{Y=[1]} \rangled_{N}$ \eqref{[1]} we obtain
    \bea
    \beta^{\alpha_{1} + \alpha_{2}+bN}_{\Delta_{1},\Delta_{2}} ([1])\Big|_{{\rm free}}
     =     \frac{\Delta + \Delta_{1} - \Delta_{2}}{2 \Delta}\Big|_{\alpha = \alpha_{1}+ \alpha_{2}+bN}.
    \eea
  This agrees with \eqref{beta1} with $\Delta_{3} \rightarrow \Delta$.
  The strategy to compute the higher order terms is the following:
  rewrite $x^{Y}$ in terms of the Jack polynomial $P_{W}(x)$ which is specified again by the partition $W$.
  (See appendix \ref{sec:Selberg} for detail.)
  By writing $x^{Y} = \sum_{W} P_{W}(x) C_{Y,W}$, we have
    \bea
    \beta^{\Delta}_{\Delta_{1}, \Delta_{2}}(Y) \Big|_{{\rm free}}
     =     \sum_{Y',W, |Y'| \leq |Y|} H_{Y,Y'} C_{Y',W} \frac{\langled P_{W}(x) \rangled_{N}}{\langled 1\rangled_{N}}.
    \eea
  The right hand side can be calculated by performing the integration 
  $\langled P_{W}(x) \rangled_{\alpha_{1},\alpha_{2},b}$ \eqref{MK}.
  The equivalence with the Virasoro three-point function was checked in lower levels in \cite{Mironov:2010ym}.
  
  Note that this equivalence is only valid for an integer $N$.
  However, the result is a rational function of $N$.
  Therefore we analytically continue $N$ to an arbitrary complex number.

\paragraph{Four-point conformal block}
  Now let us compute the partition function.
  We will below focus on the matrix model with $n=4$ which corresponds to a sphere with four punctures.
  In this case we define
    \bea
    \hat{Z}
     =     C(q) \left( \prod_{I=1}^{N_{1}} \int_{0}^{q} d \lambda_I \right)
           \left( \prod_{I=N_{1}+1}^{N} \int_{1}^{\infty} d \lambda_I \right) 
           \prod_{I < J} (\lambda_I - \lambda_J)^{-2b^2} e^{-\frac{b}{g_s} \sum_I W(\lambda_I)},
           \label{finiteN}
    \eea
  where $C(q) = q^{-2 \alpha_{0} \alpha_{2}} (1-q)^{-2\alpha_{1}\alpha_{2}}$.
  As proposed in \eqref{momcon}, the internal momentum $\alpha$ is given by
    \bea
    \alpha
     =     \alpha_{0} + \alpha_{2} + b N_{1}
     =   - \alpha_{1} - \alpha_{3} - b N_{2} + Q.
           \label{vertexmom}
    \eea
  The above prescription of the contour and the relation between $N_{1},N_{2}$ and the external momenta 
  was first given in \cite{Mironov:2010zs} (see \cite{Mironov:2009ib}) and elaborated in 
  \cite{Itoyama:2010ki,Mironov:2010ym}.
  (We are following the choice of the integration contours in \cite{Itoyama:2010ki}.)
  This integral can be expanded in $q$
    \bea
    \hat{Z}
     =     Z_{0} J, ~~~~~~
    J
     =     \sum_{k=0}^{\infty} J_{k} q^{k},
           \label{ZZZ}
    \eea 
  where $J_{k}$ are normalized such that $J_{0} = 1$, $Z_{0} = c q^{\delta}$, $\delta$ is a function of the conformal dimensions, 
  and $c$ is an irrelevant factor.
  The proposal of the equivalence between the integral representation and the conformal block is thus
    \bea
    J_{k}
     =     \CB_{k}.
    \eea
  Let us check this below. 
  
  For convenience, we change the variables as
    \bea
    \lambda_{I} 
     =     \left\{ \begin{array}{ll}
           q x_{I} & I=1,\ldots,N_{1} \\
           1/y_{I-N_{1}} & I=N_{1}+1,\ldots, N_{1}+N_{2} \\
           \end{array} \right.
    \eea
  by which the partition function becomes
    \bea
    \hat{Z}
    &=&    C'(q) \prod_{I=1}^{N_{1}} \int_{0}^{1} d x_{I} \prod_{I=1}^{N_{1}} x_{I}^{-2b\alpha_{0}}
           (1-x_{I})^{-2b\alpha_{2}} (1 - q x_{I})^{-2b\alpha_{1}} \prod_{1\leq I <J\leq N_{1}} (x_{I} - x_{J})^{-2b^{2}}
           \nonumber \\
    & &    ~~~~~~\prod_{I=1}^{N_{2}} \int_{0}^{1} d y_{I} \prod_{I=1}^{N_{2}} y_{I}^{-2b\alpha_{3}}
           (1-y_{I})^{-2b \alpha_{1}} (1 - q y_{I})^{-2b\alpha_{2}} \prod_{1\leq I <J\leq N_{2}} (y_{I} - y_{J})^{-2b^{2}}
           \nonumber \\
    & &    ~~~~~~\prod_{I=1}^{N_{1}} \prod_{J=1}^{N_{2}} (1 - q x_{I} y_{J})^{-2b^{2}},
    \eea
  where $C'(q) = q^{\Delta - \Delta_{0} - \Delta_{2}} (1-q)^{-2\alpha_{1}\alpha_{2}}$.
  This can be thought of as the double Selberg-type integral.
  By defining $\langled \ldots \rangled_{N_{1},N_{2}}$ as the average of the double Selberg integral, 
  the partition function is written as
    \bea
    \hat{Z}
     =     C'(q) \langled 1 \rangled_{N_{1},N_{2}} \langled \prod_{I=1}^{N_{1}} \prod_{J=1}^{N_{2}}
    (1 - q x_{I})^{-2b\alpha_{1}}(1 - q y_{J})^{-2b\alpha_{2}}(1 - q x_{I} y_{J})^{-2b^{2}} \rangled_{N_{1},N_{2}}.
    \eea 
  Therefore, we obtained $c = \langled 1 \rangled_{N_{1},N_{2}}$, $\delta = \Delta - \Delta_{0} - \Delta_{2}$ 
  and \cite{Itoyama:2010ki}
    \bea
    J 
    &=&    (1-q)^{-2\alpha_{1}\alpha_{2}} \langled \prod_{I=1}^{N_{1}} \prod_{J=1}^{N_{2}}
           (1 - q x_{I})^{-2b\alpha_{1}}(1 - q y_{J})^{-2b\alpha_{2}}(1 - q x_{I} y_{J})^{-2b^{2}} \rangled_{N_{1},N_{2}}
           \nonumber \\
    &=&     \langled \exp \left( 2\sum_{k=1}^{\infty}
            \frac{q^{k}}{k} \left( b\sum_{I}x_{I}^{k} + \alpha_{2} \right)\left( b\sum_{J}y_{J}^{k} + \alpha_{1} \right) 
            \right) \rangled_{N_{1},N_{2}}.
            \label{J}
    \eea
  For example, the first order term in $q$ is given by 
    \bea
    2 \langled \left( b\sum_{I}x_{I}^{k} + \alpha_{2} \right)\left( b\sum_{J}y_{J}^{k} + \alpha_{1} \right) \rangled_{N_{1},N_{2}}
     =     \frac{(\Delta + \Delta_0 - \Delta_2)(\Delta + \Delta_1 - \Delta_3)}{2 \Delta},
    \eea
  by using the formulas of the Selberg integral, with the identification \eqref{vertexmom}.
  This is indeed the conformal block at the level $1$, $\CB_{1}$.
  In principle it is possible to compute higher order terms in $q$ by using the Selberg integral formula
  and its generalization.

\paragraph{Relation to Nekrasov partition function}
  So far we focused on the relation between the integral representation and the conformal block.
  At the same time, one can argue the relation to the Nekrasov partition function
  as considered in \cite{Morozov:2013rma} following \cite{Itoyama:2010ki,Mironov:2010pi}.
  To do that we use the following expression of $J$ instead of \eqref{J}:
    \bea
    J
    &=&    (1-q)^{-2\alpha_{1}\alpha_{2}}
            \\
    & &    \langled \exp \left(b^{2} \sum_{k=1}^{\infty}
            \frac{q^{k}}{k} \left( \left( \sum_{I}x_{I}^{k} + \frac{2\alpha_{2}}{b} \right) \sum_{J}y_{J}^{k}
         +  \sum_{I}x_{I}^{k}  \left( \sum_{J}y_{J}^{k} + \frac{2\alpha_{1}}{b} \right) \right)
            \right) \rangled_{N_{1},N_{2}}
            \nonumber
    \eea
  At this stage we note that the pre-factor $(1-q)^{-2\alpha_{1} \alpha_{2}}$ is the inverse of the $U(1)$ factor
  introduced in section \ref{subsec:DF}.
  Therefore from \eqref{Nek} the second line is conjectured to be identified with the Nekrasov partition function.
  Indeed the second line has a form of summing over two Young diagrams, $\mu$ and $\nu$:
    \bea
    \sum_{\mu, \nu} q^{|\mu|+|\nu|} Z_{\mu, \nu}
    \eea
  where $Z_{\mu, \nu}$ is a double Selberg average of polynomials specified by $\mu$ and $\nu$.
  In \cite{Morozov:2013rma}, it was found that by using a particular generalization 
  of the Jack polynomial which depends on a pair of Young diagram,
  $Z_{\mu, \nu}$ is identified with the corresponding Nekrasov partition function $Z_{{\rm Nek} \mu, \nu}$
  for given $\mu$ and $\nu$.
  While the Selberg average of the generalization of the Jack polynomial is not completely understood, 
  this is a profound way towards showing the AGT correspondence.
  
  A similar calculation has been done in \cite{Zhang:2011au,Mironov:2013oaa} for the $A$-type quiver matrix model
  in section \ref{subsec:quiver}
  by making use of the Selberg integral to see the relation with four-dimensional $SU(N)$ gauge theory.
  This method has also been performed in the generalized matrix model for the one-punctured torus 
  presented in section \ref{subsec:torus} in \cite{Mironov:2010su}, 
  and the partition function has been checked to agree with the Virasoro conformal block on the torus
  in the expansion in the complex structure. 
  
\section{Conclusion and discussion}
\label{sec:conclusion}
  We have reviewed the $\beta$-deformed matrix model associated to the conformal block of two-dimensional CFT
  and instanton partition function of four-dimensional $\CN=2$ gauge theory, introduced in \cite{Dijkgraaf:2009pc}.
  This matrix model is originally motivated from the topological string theory, and
  this interesting part will be seen in the accompanying review \cite{A} in this volume.
  
  It would be interesting to consider the $\beta$-deformed matrix model corresponding to 
  asymptotically free $\CN=2$ gauge theory.
  Such models were found first in \cite{Eguchi:2009gf} for $SU(2)$ theory with $N_{f}=2,3$ hypermultiplets
  and in \cite{Nishinaka:2012kn,Rim:2012tf,Choi:2013caa} for $SU(2)$ theory coupled to superconformal field theory 
  of Argyres-Douglas type,
  which are related with irregular conformal blocks 
  in the CFT \cite{Gaiotto:2009ma,Marshakov:2009gn,Bonelli:2011aa,Gaiotto:2012sf}.
  The former model was elaborated in \cite{Itoyama:2010na} by calculating directly the integral 
  as in section \ref{sec:finiteN} and in \cite{Fujita:2009gf,Krefl:2012jt} 
  by using the loop equation to see the agreement with the subleading expansion in $\epsilon_{1,2}$.
  It was also found in \cite{Mironov:2010xs} 
  the matrix model corresponding to the $SU(2)$ super Yang-Mills theory.
  
  Another interesting generalization is
  the $q$-deformed matrix model related to the Nekrasov partition function of the five-dimensional gauge theory 
  proposed in \cite{Schiappa:2009cc,Awata:2010yy,Mironov:2011dk}.
  It would be interesting to elaborate this model further in the context of topological string theory.
  
  In \cite{Klemm:2008yu,Sulkowski:2009ne,Marshakov:2011vw,Kimura:2011zf} a different matrix model 
  which describes the Nekrasov partition function has been found.
  While the form of the potential in particular is quite different,
  it would be interesting to see the relation with the model in this review.
  

\section*{Acknowledgements}
The author would like to thank Giulio Bonelli, Tohru Eguchi, Hiroshi Itoyama, Takeshi Oota, Alessandro Tanzini,
and Futoshi Yagi
for stimulating collaborations on the $\beta$ deformed matrix model.
The author would like to thank Kazuo Hosomichi, Peter Koroteev and Pavel Putrov for helpful discussions and useful comments.
The work of the author is supported by the EPSRC programme grant ``New Geometric
Structures from String Theory'', EP/K034456/1.


\appendix

\section*{Appendix}

\section{Integral formulas}
\label{sec:Selberg}
  Let us define the following multiple integral
    \bea
    \langled x^{Y}\rangled_{N}
    &=&    \prod_{I=1}^{N} \int_{0}^{1} d x_{I} \prod_{I=1}^{N} x_{I}^{\alpha}
           (1-x_{I})^{\beta} \prod_{1\leq I <J\leq N} (x_{I} - x_{J})^{2\gamma} x^{Y}
    \eea
  where 
  supposing that $\Re \beta >0$, ... for convergence of the integrals.
  When $Y=\emptyset$ and $Y = [1^{k}]$, this is the Selberg integral \cite{Selberg} and Aomoto integral \cite{Aomoto}
    \bea
    \langled 1\rangled_{N}
    &=&    \prod_{j=0}^{N-1}\frac{\Gamma(\alpha + 1 + j\gamma) \Gamma(\beta + 1 + j \gamma) 
     \Gamma(1 + (j+1)\gamma)}{\Gamma(\alpha + \beta + 2 + (N + j-1)\gamma) \Gamma(1 + \gamma)},
           \nonumber \\
    \langled x^{Y=[1^{k}]}\rangled_{N}
    &=&    \langled 1\rangled_{N} \prod_{j=1}^{k} \frac{\alpha + 1 +(N-j)\gamma}{\alpha + \beta + 2 + (2N - j -1)\gamma}.
           \label{[1]}
    \eea
  
  Another multiple integral which appeared in the main text is involving the Jack polynomial $P_{Y}(x)$.
  This is a polynomial of $(x_{1},x_{2},\ldots,x_{N})$ and written as
    \bea
    P_{Y}(x)
     =     m_{Y}(x) + \sum_{Y'<Y} a_{Y,Y'}m_{Y'}(x),
    \eea 
  where $m_{Y}(x)$ is the monomial symmetric polynomial.
  Then the following integral is given by \cite{Kaneko,Kadell}
    \bea
    \langled P_{Y}(x) \rangled_{N}
    &=&    \prod_{I=1}^{N} \int_{0}^{1} d x_{I} \prod_{I=1}^{N} x_{I}^{\alpha}
           (1-x_{I})^{\beta} \prod_{1\leq I <J\leq N} (x_{I} - x_{J})^{2\gamma} P_{Y}(x)
           \label{MK} \\
    &=&    \prod_{i\geq1}\prod_{j=0}^{y_{i}-1} \frac{\alpha + 1 +j+(N-i)\gamma}{\alpha + \beta + 2 + j+ (2N - i -1)\gamma}
           \frac{\prod_{i\geq1}\prod_{j=0}^{y_{i}-1} (N+1-i)\gamma + j}{\prod_{(i,j)\in Y} (y_{i}-j+(\tilde{y}_{j} - 
           i+1)\gamma)},
           \nonumber 
    \eea
  where $Y=[y_{1},y_{2},\ldots]$ with $y_{1}\geq y_{2}\geq \ldots$ 
  and $\tilde{Y} = [\tilde{y}_{1} \geq \tilde{y}_{2} \geq \ldots]$ is the transpose of $Y$.


\paragraph{\large References to articles in this volume}
\renewcommand{\refname}{

\paragraph{Other references}
\renewcommand{\refname}{\vskip-36pt}

\bibliographystyle{ytphys}
\small\baselineskip=.97\baselineskip
\bibliography{ref}

\end{document}